\documentclass[pdflatex,sn-mathphys-num]{sn-jnl}

\usepackage{xcolor} 
\usepackage{algorithm,algorithmic} 
\usepackage{amsthm, amssymb,amsmath,amsfonts} 
\usepackage{graphicx}
\usepackage{subcaption}
\usepackage{tabularx} 
\usepackage{diagbox}
\usepackage{multirow}
\usepackage{comment}
\usepackage{bm}
\usepackage{bbm}
\usepackage{xcolor}
\usepackage{enumitem}
\floatname{algorithm}{Algorithm}

\newcommand{\norm}[1]{\left\lVert#1\right\rVert}

\newtheorem{remark}{\textbf{Remark}}[section]
\newtheorem{defi}{\textbf{Definition}}[section]  
\newtheorem{asmp}{\textbf{Assumption}}[section]  
\newtheorem{thm}{\textbf{Theorem}}[section]

\usepackage[english]{babel}
\usepackage[utf8]{inputenc}
\usepackage[T1]{fontenc}
\usepackage{helvet}
\usepackage{etoolbox}
\usepackage{graphicx}
\usepackage{titlesec}
\usepackage{caption}
\usepackage{booktabs}
\usepackage{fancyhdr}
\usepackage{graphicx}
\usepackage{natbib} 
\usepackage{hyperref}

\begin{document}

\title[Article Title]{Piecewise Deterministic Sampling for Constrained Distributions}

\author*[1,2]{\fnm{Jo\"{e}l} \sur{Tatang Demano}}\email{id2005@hw.ac.uk}

\author[2]{\fnm{Paul} \sur{Dobson}}\email{p.dobson$_{-}$1@hw.ac.uk}
\equalcont{All the authors contributed equally to this work.}

\author[3]{\fnm{Konstantinos} \sur{Zygalakis}}\email{k.zygalakis@ed.ac.uk}
\equalcont{All the authors contributed equally to this work.}

\affil*[1]{\orgdiv{}\orgname{Maxwell Institute for Mathematical Sciences}, \orgaddress{\street{Bayes Centre}, \city{} \postcode{47 Potterrow EH8 9BT}, \state{} \country{Scotland}}} 

\affil[2]{\orgdiv{Department of Mathematics and Computer Sciences}, \orgname{Heriot-Watt University}, \orgaddress{\street{}
\city{Edinburgh}, \postcode{EH14 4AS}, \state{} \country{Scotland}}}

\affil[3]{\orgdiv{School of Mathematics}, \orgname{University of Edinburgh}, \orgaddress{\street{} \city{Edinburgh}, \postcode{EH9 3FD}, \state{} \country{Scotland}}}


\abstract{In this paper, we propose a novel class of Piecewise Deterministic Markov Processes (PDMPs) that are designed to sample from 
probability distributions $\pi$ supported on a convex set $\mathcal{M}$. This class of PDMPs adapts the concept of a mirror map from convex optimisation to address sampling problems.  The corresponding algorithms provide unbiased samples that respect the constraints and, moreover, allow for exact subsampling. We demonstrate the advantages of these algorithms against a range of constrained sampling problems where the proposed algorithms outperform  state of the art stochastic differential equation-based methods.
}




\maketitle


\section{Introduction}
The ability to generate random variables according to a given probability distribution with density of the form
\begin{equation} \label{eq:prob}
    \pi(x) \propto e^{-U(x)}, \quad x\in \mathcal{M}
\end{equation}
 is central to Bayesian computation, and arises in a number of applications such as image processing \cite{green2015bayesian}, statistical mechanics \cite{Leimkuhler15}, and inverse problems \cite{Stuart_2010}.  Here $U:\mathcal{M}\to\mathbb{R}$ is a given function,  and $\mathcal{M}$ is a convex subset of $\mathbb{R}^d$. Markov chain Monte Carlo (MCMC) methods deal with this problem by constructing a Markov chain that is stationary with respect to $\pi$. Two widely used approaches to construct such a Markov chain are as follows: 1) Based on an accept-reject mechanism \citep{metropolis1953equation}; 2) Based on stochastic differential equations (SDEs) such as the Langevin SDE \citep{Roberts96}. Algorithms based on the first approach may suffer from poor mixing properties, especially in high dimensions \citep{DM19}. On the other hand, SDE-based algorithms can exhibit faster convergence properties at the cost of some (controlled) bias \citep{durmus2017nonasymptotic,JMLR:v22:21-0453,Dalalyan19}. 
 
Piecewise Deterministic Markov Processes (PDMPs) were introduced in the early 1980s by \cite{davis1984piecewise} in the context of biological and communication systems. They are non-diffusive, irreversible Markov processes that evolve deterministically between random times. Similarly to the case of SDE samplers, it is possible to design PDMPs that are invariant with respect to $\pi$. This has led to variety of sampling algorithms such as the Zig-Zag sampler (ZZS), \cite{bierkens2019ergodicity}, the Bouncy Particle Sampler (BPS), \cite{bouchard2018bouncy}, the Boomerang Sampler (BS), \cite{bierkens2020boomerang} and the Coordinate Sampler(CS), \cite{wu2020coordinate} to name a few. Implementation of these algorithms is discussed in \cite{fearnhead2018piecewise}. 

PDMPs are particularly attractive for MCMC due to three properties:  i) they can be simulated exactly under suitable conditions.\footnote{A detailed discussion about challenges with exact simulation is given in Section~\ref{subsec:implementation}.}; ii) they are irreversible, thus potentially converging faster towards $\pi$ \citep{diaconis2000analysis, bierkens2016non},  iii) they allow for unbiased estimation using stochastic gradients. This is useful in the context of large data sets where the cost of the algorithm can be reduced to $\mathcal{O}(1)$ cost per iteration as shown in \cite{bierkens2019zig}. 

All the algorithms mentioned above deal with unconstrained sampling problems, that is $\mathcal{M}=\mathbb{R}^d$. When $\mathcal{M}$ is a proper convex subset of $\mathbb{R}^d$, there are several approaches for SDE-based samplers that borrow ideas from convex optimisation. In particular, a natural extension of the projected gradient descent \citep{Beck17}, the projected Langevin Monte Carlo method \cite{NIPS2015_c0f168ce} incorporates a projection back to $\mathcal{M}$ after every Euler-Maruyama step. Alternatively, one can use a regularised version of the indicator function $\bm{1}_\mathcal{M}$ to define a probability density with support over $\mathbb{R}^{d}$ and employ different numerical discretizations of the underlying SDE leading for example to the Moreau Yosida Unadjusted Langevin Algorithm (MYULA) \cite{durmus2018efficient,brosse2017sampling} or the Forward Backward Unadjusted Langevin algorithm (FBULA) \cite{eftekhari2023forward}.  The mirror Langevin algorithms use the mirror map to ensure that the samples produced reside in $ \mathcal{M} $ \citep{hsieh2018mirrored, zhang20a, Chewi20}. Finally, \citep{GC11,BLS25} add a Riemannian metric to the space $\mathcal{M}$ and use suitable discretizations of geometrically inspired SDEs.

Ideally, one would like to combine the ability of the SDE-based samplers mentioned above to deal with the fact that $\mathcal{M}$ is a proper convex subset of $\mathbb{R}^{d}$ with the desirable features of PDMPs, namely unbiasedness, and the ability to subsample. One approach \cite{bierkens2023methods} to do this is to view the probability distribution $\pi$ as a discontinuous function of $\mathbb{R}^d$ and then add reflections to standard PDMPs to construct a PDMP that respects the constraints. In this paper we take a different approach and  
introduce a framework that adapts PDMPs to include mirror maps and thus allows sampling in the case when $\mathcal{M}$ is a proper convex subset of $\mathbb{R}^{d}$. In addition, we specify a more detailed choice of PDMP called mirror Zig-Zag sampler (MZZS) and establish conditions for its ergodicity. 

The remainder of this paper is organized as follows: in Section \ref{sec2}, we discuss in detail SDE-based samplers both in the constrained and the unconstrained cases. Furthermore, in Section \ref{sec3} we give a broad introduction to PDMPs in the unconstrained setting. Section
\ref{mirror-PDMPs} extends these ideas to the constrained setting and explains the role of the mirror map for PDMPs, while Section \ref{sec4} shows a number of numerical experiments that illustrate the benefits of mirror PDMPs.


\section{SDE-based samplers}
\label{sec2}
In this section, we  review some standard approaches for sampling from a density of the form \eqref{eq:prob} based on SDEs. We  distinguish two cases, the first being when $\mathcal{M}=\mathbb{R}^{d}$ and the second when $\mathcal{M}$ is a proper convex subset of $\mathbb{R}^{d}$.
\subsection{The unconstrained case}
There exist several different SDEs that have $\pi$ as their invariant distribution. As shown in \cite{Fox15} these SDEs can be summarised by the following family
\begin{eqnarray}\label{eq:SDE_general_form}
    dX_{t} &=& [-(S(X_{t})+J(X_{t}))\nabla U(X_{t})+\Gamma(X_{t})]dt+\sqrt{2S(X_{t})}dW_{t} \nonumber \\
\Gamma_{i}(x) &=&\sum_{j=1}^{n} \partial_{x_{j}}(S_{ij}(x)+J_{ij}(x))
\end{eqnarray}
for some semi-definite diffusion matrix $S(x)$ and skew-symmetric matrix $J(x)$. Here $(W_{t})_{t \geq 0}$ is a standard $d$-dimensional Brownian motion.

Different choices of matrices $S(x), J(x)$ lead to different SDEs. The simplest one is the Langevin SDE corresponding to the choice $S(x)=I_{d}, J(x)=0$
\begin{equation} \label{eq:overd}
dX_{t}=-\nabla U(X_{t})dt+\sqrt{2}dW_{t}.
\end{equation}
The simplest method for solving \eqref{eq:overd} is the Euler-Maruyama method \citep{KP92}
\begin{equation} \label{eq:EM}
\bar{X}_{n+1}= \bar{X}_{n}-\Delta t \nabla U(\bar{X}_{n}) +\sqrt{2\Delta t} \xi_{n},    
\end{equation}
where $\Delta t >0$ is the time-step and $(\xi_{n})_n$ is sequence of i.i.d standard Gaussian random variables on $\mathbb{R}^{d}$.
\subsection{The constrained case}
There are several different approaches for dealing with the case when $\mathcal{M}$ is a proper convex subset of $\mathbb{R}^{d}$. One such approach is based on the reflected SDE (RSDE)
\begin{equation*}
    dX_t = -\nabla U(X_t)dt  +\sqrt{2}dW_t - n(X_t)dL_t,
\end{equation*}
where $L_t$ is the boundary local time (a non decreasing process that increases only when $X_t\in \partial\mathcal{M}$) and $n$ is the outward unit normal vector field. There is a variety of discretisation methods applicable to this RSDE see for example  \cite{cattiaux2017invariant,Bossy04,leimkuhler2020simplest,pettersson1997penalization,Gobet2001}. Here, we will concentrate on two discretisations: the projected Langevin Monte Carlo method (PLMC) \cite{NIPS2015_c0f168ce} and the penalty method which corresponds to MYULA for this RSDE. The projected Langevin Monte Carlo method (PLMC) given by
\begin{equation} \label{eq:PEM}
\bar{X}_{n+1}=P_{\mathcal{M}}(\bar{X}_{n}-\Delta t \nabla U(\bar{X}_{n}) +\sqrt{2\Delta t} \xi_{n}) 
\end{equation}
adds a projection $P_{\mathcal{M}}$ back to $\mathcal{M}$ after each
Euler-Maruyama step \eqref{eq:EM} to ensure that iterates of \eqref{eq:PEM} remain within $\mathcal{M}$. The penalty method \cite{pettersson1997penalization} uses a regularised version of the indicator function $\bm{1}_{\mathcal{M}}$ to construct the potential
\begin{equation}
U^{\varepsilon}(x)=U(x)+ \frac{1}{2\varepsilon}\|{x-P_{\mathcal{M}}(x)}\|^{2},    
\end{equation}
which now defines a probability density with support over the whole of $\mathbb{R}^{d}$ where $\|\cdot\|$ denotes the Euclidean norm. Discretizing, the Langevin equation \eqref{eq:overd} with the Euler-Maruyama method for the potential $U^{\varepsilon}$ gives 
\begin{equation} \label{eq:MYULA}
\bar{X}_{n+1}=\bar{X}_{n}-\Delta t \nabla U(\bar{X}_{n})+\frac{\Delta t}{\varepsilon}\left(P_{\mathcal{M}}(\bar{X}_{n})-\bar{X}_{n} \right)+\sqrt{2\Delta t} \xi_{n}.
\end{equation}
One drawback of \eqref{eq:MYULA} relates to the fact that one needs to operate with a small value of the regularisation parameter $\varepsilon$. This results in $\Delta t$ being at most $\mathcal{O}(\varepsilon)$, which in turn leads to slow mixing. On the other hand \eqref{eq:PEM} does not impose an additional constraint on $\Delta t$ and can be shown to be ergodic for any value of $\Delta t$ provided $\mathcal{M}$ is bounded and connected. However, using large values of $\Delta t$ leads to a large bias. 

Note there are other approaches such as those based on reflective Hamiltonian Monte Carlo \cite{chalkis2021truncated} which consider the case where $\mathcal{M}$ is a convex polytope and Wall Hamiltonian Monte Carlo \cite{pakman2014exact} which is specialised to the case where $U$ is a quadratic function.

\subsubsection{SDEs exploiting the mirror map}
An alternative way to deal with constraints similar to those already discussed is by using a
mirror map.  In the case of convex optimization 
\citep{Beck17} mirror descent
\begin{subequations}\label{eq:md}
\begin{align}
\bar{\zeta}_{n+1} &=\bar{\zeta}_{n}-\Delta t \nabla U(\bar{X}_{n}), \\
\bar{X}_{n+1}&=\nabla \psi^{*}(\bar{\zeta}_{n+1}),
\end{align}
\end{subequations}
improves on projected gradient descent. Here $\psi$ is a barrier function defined as follows:
\begin{defi}\label{Mirror-Map-def}
A barrier function associated to the set $\mathcal{M}$ is a map $\psi: \text{int}(\mathcal{M}) \to \mathbb{R}$ which is, proper, closed, strictly convex, twice continuously differentiable and such that 
\begin{equation*}
\|\nabla \psi({x_n})\| \to \infty \quad \text{as} \ n \to \infty, 
\end{equation*} 
whenever  $x_1,x_2,\ldots$ is a sequence in $\mathcal{M}$ converging to $ x \in \partial \mathcal{M}$
where $\partial \mathcal{M}$ denotes the boundary of $\mathcal{M}$ and $\text{int}(\mathcal{M})$ its interior. 
\end{defi}
Some examples of commonly used barrier functions \cite{hsieh2018mirrored, zhang20a,srinivasan2024fast} along with the corresponding set $\mathcal{M}$ are presented in Table \ref{tab:barriers}.
\begin{table}[h!]
    \centering
    \begin{tabular}{|c|c|}
    \hline
     \text{Domain $\mathcal{M}$}    &  \text{Barrier function} \\
     \hline
     \text{Rectangle:} $\mathcal{M}=\{x\in \mathbb{R}^{d}: a_{i} \leq x_{i} \leq b_{i} \}$ &
      $\psi(x) = -\sum_{i=1}^{d} \log(x_i-a_i)+ \log(b_i-x_i)$, \\
     \hline 
      \text{Euclidean ball:} $\mathcal{M} = \{ x \in \mathbb{R}^d: \|x\| \leq 1 \}$   &  $\psi(x) = -\log(1-\|x\|)- \|x\|$  \\
      \hline 
      $\text{Polytope}(A,b): \mathcal{M} = \{x \in \mathbb{R}^d: Ax \leq b \}, \ A \in \mathbb{R}^{m \times d}, \ b \in \mathbb{R}^m$ & $\psi(x) = -\sum_{i=1}^m \log(b_i-a_i^\top x)$ \\
      \hline 
      $\text{Lorentz cone:} \ \mathcal{M} = \{(x_1,x_2) \in \mathbb{R}^2: |x_1|< x_2\}$ & $\psi(x) = -\log(x_2^2-x_1^2)$ \\ 
      \hline 
      $\text{Positive orthant: } \ \mathcal{M} = (a_i, +\infty)^d = \{x \in \mathbb{R}^d: x_i > a_i\}$ & $\psi(x) = \frac{1}{2}\|x\|^{2} -\sum_{i=1}^d \log(x_i-a_i)$ \\
      \hline 
      $\text{Simplex}: \Delta_d = \{x \in \mathbb{R}^d: \sum_{i=1}^d x_i = 1\}, \ x_d = 1- \sum_{i=1}^{d-1}x_i$ & $\psi(x) = \sum_{i=1}^{d-1} x_i \log(x_i) + x_d \log(x_d)$ \\
      \hline
    \end{tabular}
    \caption{Examples of convex sets and their corresponding barrier functions.} 
    \label{tab:barriers}
\end{table}

 For a given barrier function $\psi$, the mirror map $\nabla \psi: \mathcal{M} \to \mathbb{R}^d$ is a bijection from the convex set $\mathcal{M}$ to $\mathbb{R}^d$ \citep[Corollary 26.3.1]{tyrrell1970convex}. Moreover, the inverse of $\nabla \psi$ is $\nabla \psi^*$ where 
\begin{equation*}
  \psi^*({y}) = \sup_{{x} \in \mathcal{M}} \quad \{{x}^\top {y} - \psi({x}) \} 
\end{equation*}
refers to the Legendre-Fenchel conjugate of $\psi$ \citep[Corollary 23.5.1]{tyrrell1970convex} in which $x^\top$ denotes the transpose of the vector $x$. We will refer to $\mathcal{M}$ as the \emph{primal} space and the image of the mirror map as the \emph{dual} space. Under the assumptions of Definition~\ref{Mirror-Map-def} the dual space is $\mathbb{R}^d$ \cite[Theorem 26.5]{tyrrell1970convex}. Note that \eqref{eq:md} can be seen as a discretisation of the mirror gradient flow  \citep{NY83}
\begin{align}\label{eq:mf}
d\zeta_{t} =-\nabla U({X}_{t}), \quad {X}_{t}=\nabla \psi^{*}(\zeta_t).
\end{align}
This equation can be seen as the natural gradient flow on the Riemannian manifold $\mathcal{M}$ with Hessian metric $\nabla^2\psi$. Due to the assumptions on the growth of $\nabla \psi$, ${X}_t$ in \eqref{eq:mf} is constrained to the set $\mathcal{M}$.

As discussed in the unconstrained case, one can write many different SDEs that have the same invariant distribution. Thus, when utilising the idea of the mirror map for SDE-based sampling, \cite{hsieh2018mirrored,zhang20a,Chewi20} proposed two different SDEs for constrained sampling. More precisely,  \cite{hsieh2018mirrored} proposed 
\begin{subequations}\label{eq:mirror1}
\begin{align} 
    X_t&=\nabla \psi^*(\zeta_t), \label{eq:mirror1x}\\
    d\zeta_t &= -\nabla V(\zeta_t) dt +\sqrt{2}dW_t, \label{eq:mirror1zeta}
    \end{align}
\end{subequations}
where $V$ is the negative log of the unnormalized density of the push-forward distribution $\tilde{\pi}=(\nabla\psi)_{\#}\pi$ and is given by
\begin{equation}\label{pushforward-potential}
	V(\zeta) = (U \circ \nabla \psi^*)(\zeta) - \log\det\nabla^2 \psi^*(\zeta).
\end{equation}
Here $\circ$ denotes function composition and $\det$ is the determinant. 
The Euler-Maruyama discretization of \eqref{eq:mirror1} is given by 
\begin{subequations}\label{eq:MLD}
\begin{align}
    \bar{\zeta}_{n+1} &= \bar{\zeta}_n -\Delta t \nabla V(\bar{\zeta}_n) + \sqrt{2\Delta t} \xi_n \label{eq:MLD_zeta},\\
    \bar{X}_{n+1} &= \nabla \psi^*(\bar{\zeta}_{n+1}), \label{eq:MLD_x}
\end{align}
\end{subequations} 
and hereafter we will refer to this as the mirror Langevin algorithm with additive noise (MLAa). A Metropolis adjusted version of this algorithm was considered and analysed in \cite{zhang20a,srinivasan2024fast}, which we will refer to as the Metropolis Adjusted Mirror Langevin Algorithm (MAMLA).

The alternative SDE 
\begin{subequations}\label{eq:mirror2}
\begin{align} 
    X_t&=\nabla \psi^*(\zeta_t),\\
    d\zeta_t &= -\nabla U(X_t) dt +\sqrt{2}[\nabla^2\psi(X_t)]^\frac{1}{2}dW_t, \label{eq:mirror2zeta}
    \end{align}
\end{subequations}
is proposed in \cite{zhang20a,Chewi20}.
An Euler discretization of \eqref{eq:mirror2} was proposed in \cite{zhang20a} and analysed in \cite{li2022mirror}, while the following more sophisticated discretization was proposed in \cite{Chewi20} 
\begin{subequations} \label{eq:MLAm}
\begin{align}
    &  \bar{X}_{n+1/2}=\nabla \psi^{*}(\nabla \psi(\bar{X}_{n})-\Delta t \nabla U(\bar{X}_{n})) \\
    & \bar{X}_{n+1} = \nabla \psi^*(Z_{\Delta t}), \qquad \text{where} \quad \begin{cases}
        dZ_t = \sqrt{2}[\nabla^2\psi^*(Z_t)]^{-1/2} dW_t,\\
        Z_0 = \nabla \psi(\bar{X}_{n+1/2}). 
    \end{cases}
    \label{eq:Z-dynamic}
\end{align}
\end{subequations}
We will refer to this discretization as the mirror Langevin algorithm with multiplicative noise (MLAm). As in \cite{Chewi20}, we discretize the $Z_t$ dynamics in \eqref{eq:Z-dynamic} using 10 steps of Euler-Maruyama algorithm for each step of MLAm.

Even though at first sight \eqref{eq:mirror1zeta} and \eqref{eq:mirror2zeta} are not the same, they both satisfy the same objective which is to construct an SDE in the unconstrained space that is ergodic with respect to the push-forward distribution $\tilde{\pi}$.
In fact, when written as an SDE purely in the primal space, they both correspond to \eqref{eq:SDE_general_form} with $J=0$ but with different choices of $S$. For \eqref{eq:mirror1} we have 
\begin{equation*}
    S(x) = [\nabla^2\psi({x})]^{-2} = [\nabla^2\psi^*(\nabla\psi({x}))]^2
\end{equation*}
whereas for \eqref{eq:mirror2} we have
\begin{equation}\label{eq:S_riemannian}
    S(x) = [\nabla^2\psi({x})]^{-1} = \nabla^2\psi^*(\nabla\psi({x})).
\end{equation}

\begin{remark}\label{rem:remannian_manifold}
    One can view the mirror map $\psi$ as imposing a Riemannian manifold structure on the set $\mathcal{M}$. The Riemannian metric on this manifold is given by $\nabla^2\psi$. With this structure one can construct the Riemannian manifold Langevin SDE as in \cite{GC11}, this is the SDE \eqref{eq:SDE_general_form} with $S$ given by \eqref{eq:S_riemannian} and hence \eqref{eq:mirror2} coincides with the Riemannian manifold Langevin SDE with the metric $\nabla^2\psi$, see \cite{zhang20a} for further details.
\end{remark}


\subsection{A numerical illustration}\label{sec:num_ill}
We  now illustrate the behaviour of all the different SDE-based samplers discussed previously, in the case where $U(x)$ is given by 
\begin{equation} \label{eq:trunc}
U(x)=\frac{1}{2}x^{\top}\Sigma^{-1}x, \quad \mathcal{M}=[-1,1]^{2}, \quad \Sigma=\begin{pmatrix}  
1 & 0 \\
0 & \epsilon
\end{pmatrix}
\end{equation} 
corresponding to the case of a truncated Gaussian. For this experiment we set $\epsilon=0.01$.  In particular, we study PLMC \eqref{eq:PEM}, MYULA \eqref{eq:MYULA}, and  MLAm \eqref{eq:MLAm}. Initially, we choose our barrier function to be
\begin{equation*} 
\psi_{1}(x) = -\frac{1}{2}\sum_{i=1}^{2} \left[\log{(x_{i}+1)}+\log{(1-x_{i})} \right]
\end{equation*}  
as it addresses the constraints associated with $\mathcal{M}$.

\begin{figure}[t]
    \centering
    \begin{subfigure}{0.19\linewidth}
        \includegraphics[width=\linewidth]{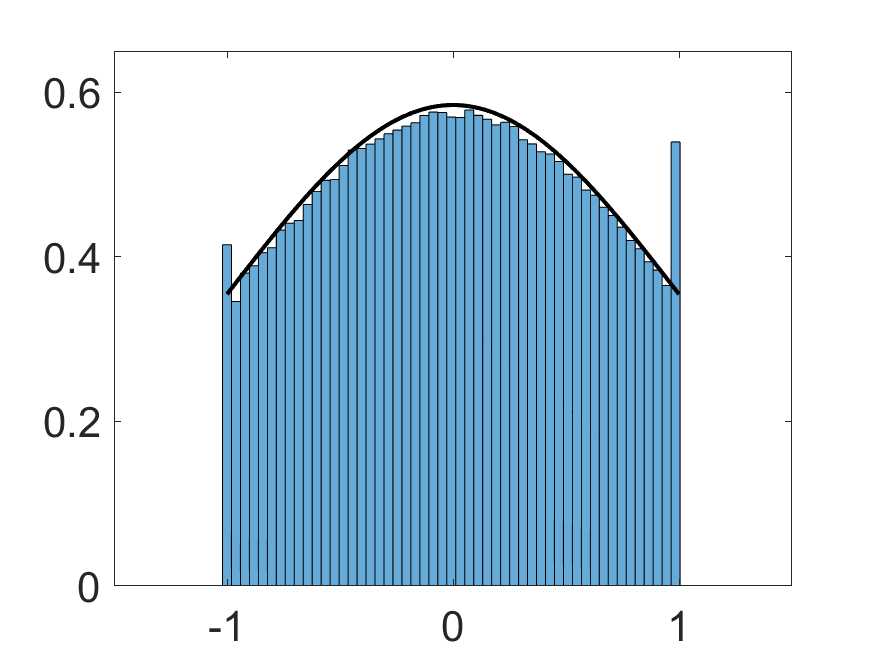}
        \caption{PLMC}
    \end{subfigure}
    \begin{subfigure}{0.19\linewidth}
        \includegraphics[width=\linewidth]{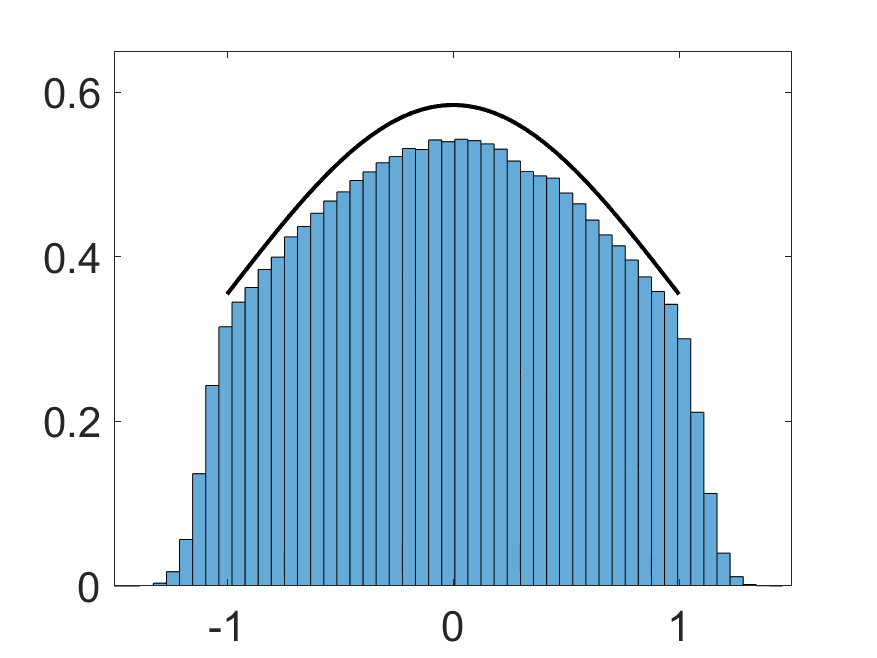}
        \caption{MYULA}
    \end{subfigure}
    \begin{subfigure}{0.19\linewidth}
        \includegraphics[width=\linewidth]{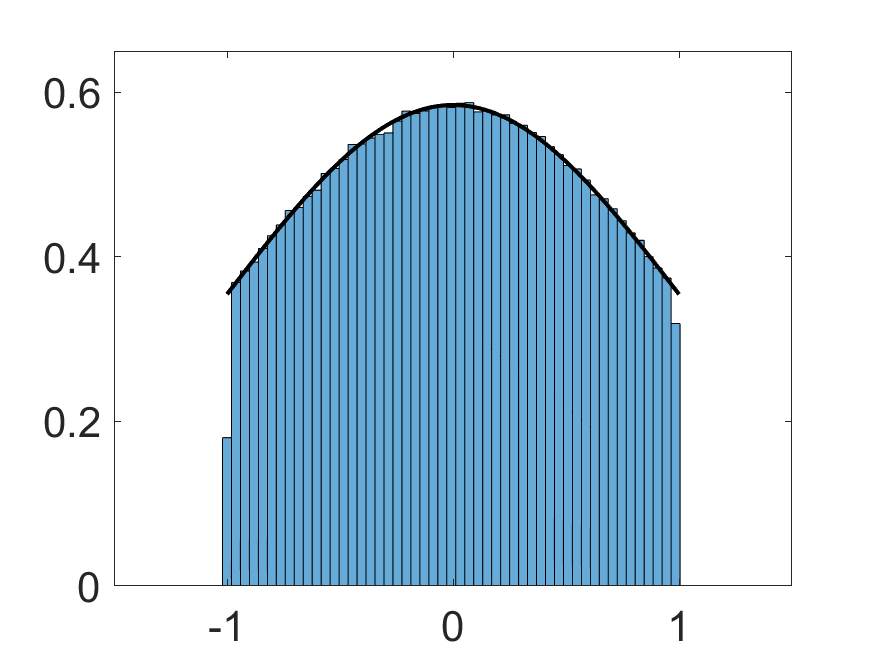}
        \caption{MLAm}
    \end{subfigure}
    \begin{subfigure}{0.19\linewidth}
        \includegraphics[width=\linewidth]{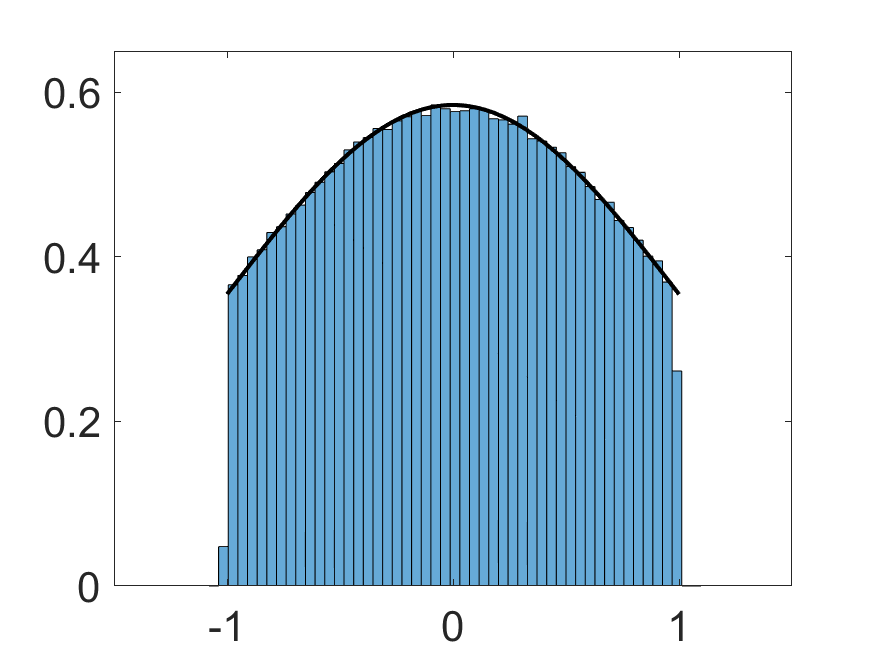}
        \caption{P-MLAm}
    \end{subfigure}
     \begin{subfigure}{0.19\linewidth}
        \includegraphics[width=\linewidth]{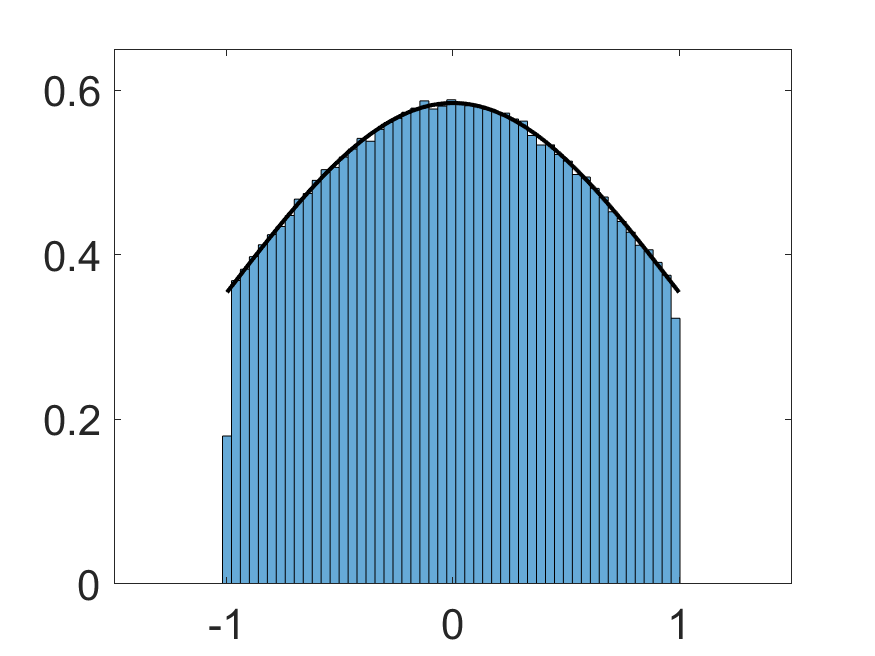}
        \caption{MAMLA} 
    \end{subfigure} \\
    \begin{subfigure}{0.19\linewidth}
        \includegraphics[width=\linewidth]{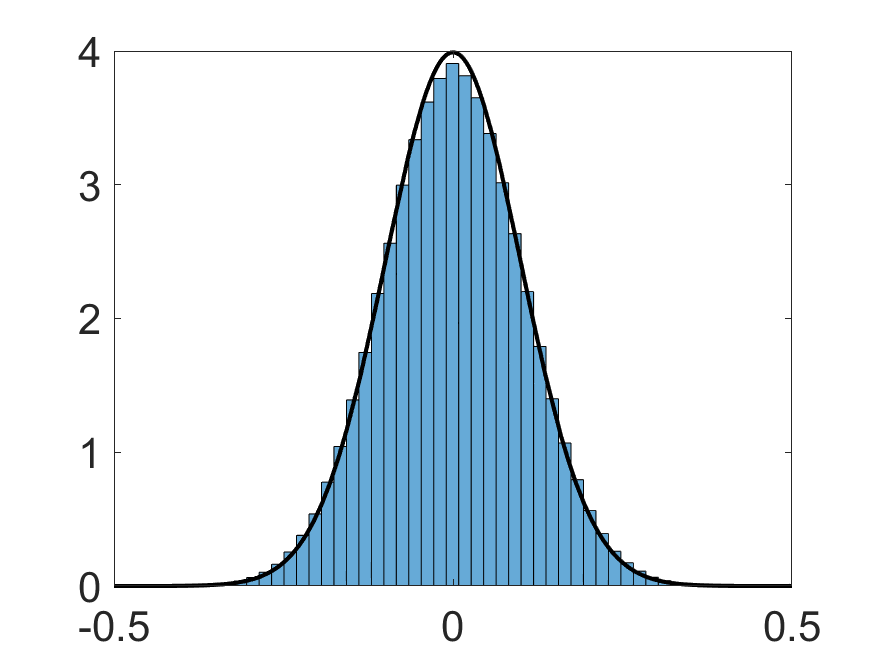}
    \end{subfigure}
    \begin{subfigure}{0.19\linewidth}
        \includegraphics[width=\linewidth]{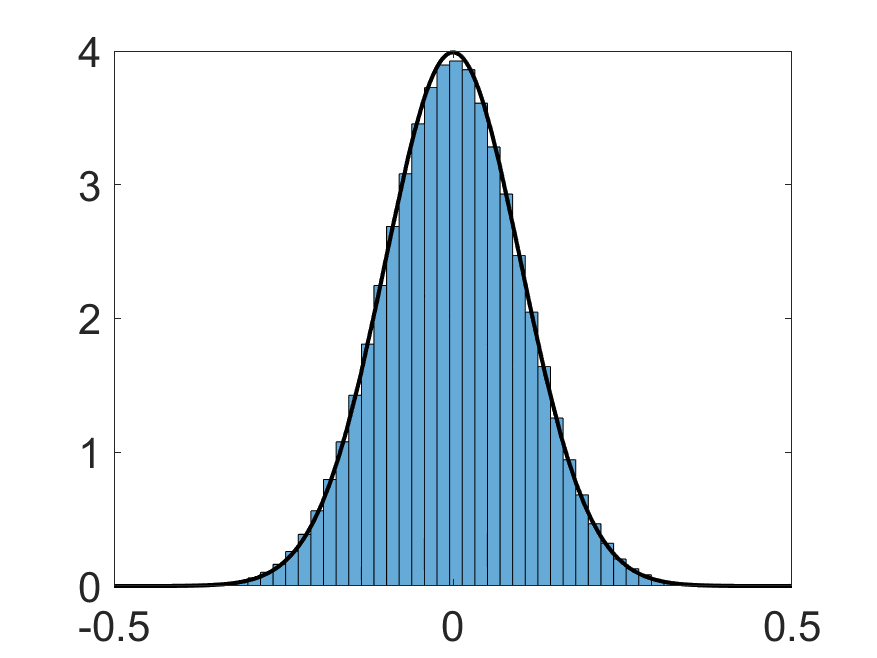}
    \end{subfigure}
    \begin{subfigure}{0.19\linewidth}
        \includegraphics[width=\linewidth]{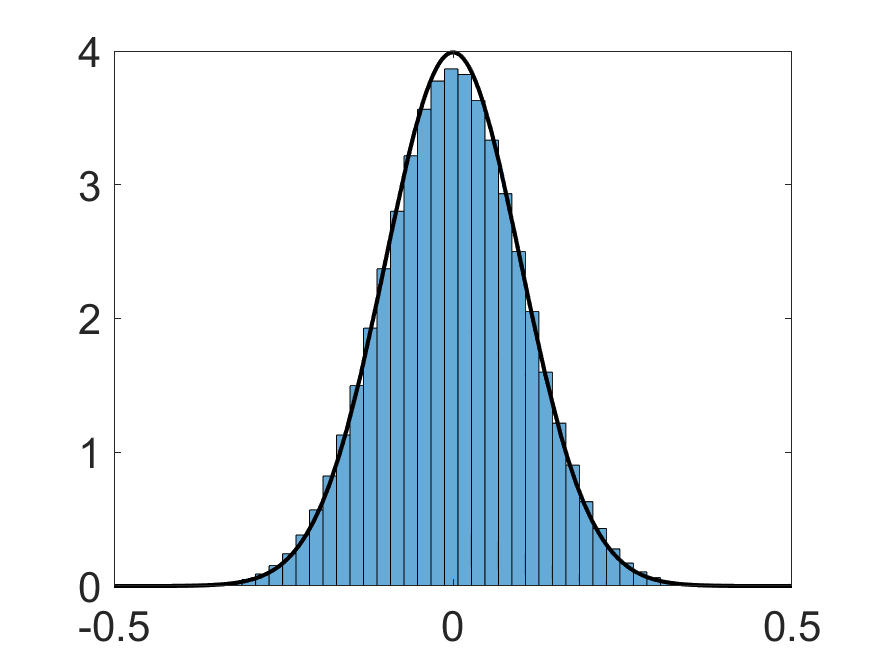}
    \end{subfigure}
    \begin{subfigure}{0.19\linewidth}
        \includegraphics[width=\linewidth]{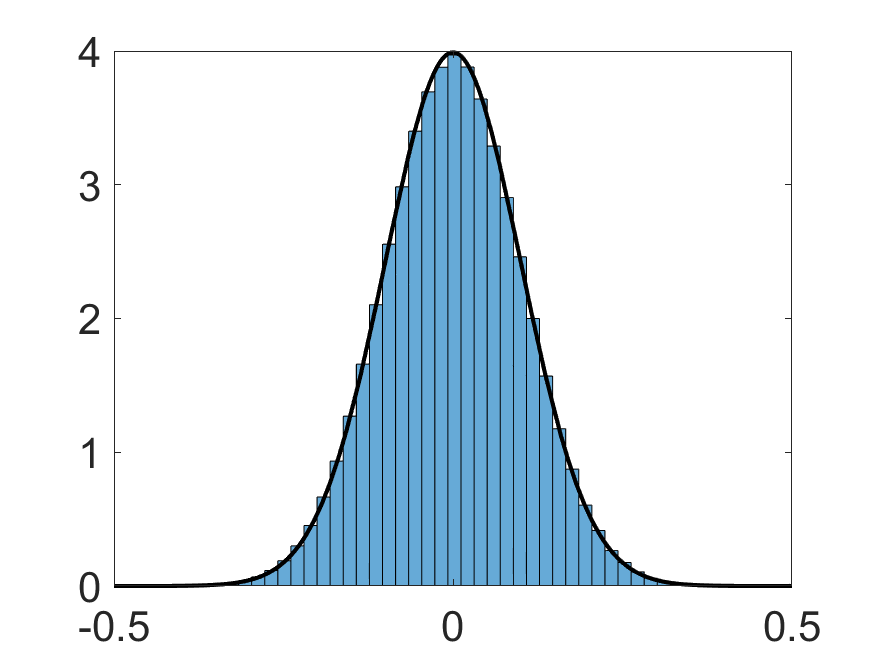}
    \end{subfigure} 
    \begin{subfigure}{0.19\linewidth}
        \includegraphics[width=\linewidth]{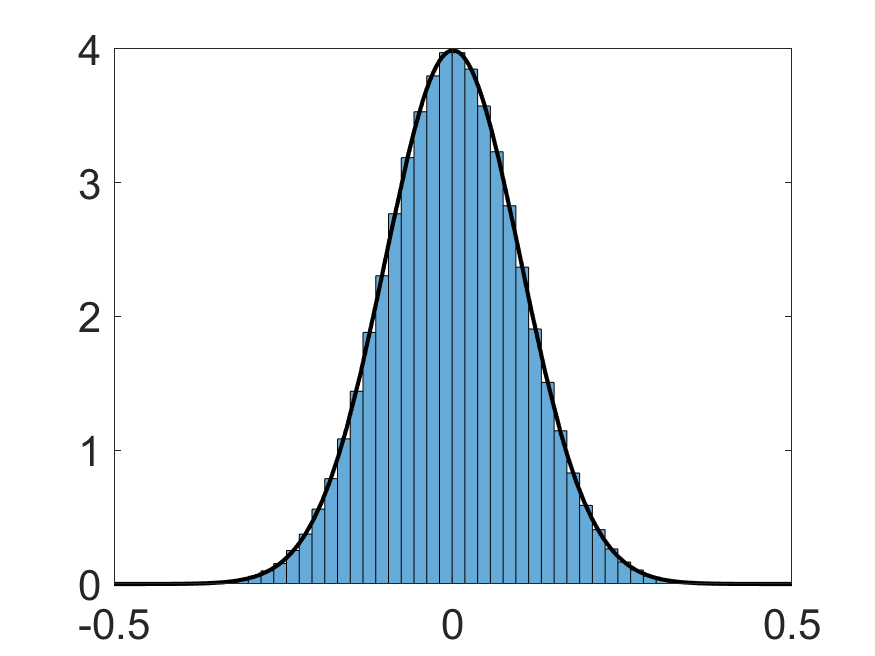}
    \end{subfigure}
    \caption{ Histograms for the two different components for the different algorithms (\textit{top}  first component and \textit{bottom}  second component) in the case of truncated Gaussian \eqref{eq:trunc}. The true density is given by the solid black line.} 
    \label{fig:histograms}
\end{figure}

We start our investigation by using $10^{6}$ realisations of each algorithm, with each realisation using $10^{4}$ gradient evaluations. We use the final sample for each of the realisations to produce a histogram for each of the methods. We plot the results in Figure \ref{fig:histograms}. As we can see, when it comes to the $x_{1}$ component both PLMC and MYULA are visually biased, with PLMC putting a lot of probability on the boundary of $\mathcal{M}$, while as expected, MYULA has a large probability of being outside $\mathcal{M}$. On the other hand, MLAm  show no visible bias.  In the case of the $x_{2}$ component there is no visible bias for any of the methods.   

To understand the speed of convergence, we calculate the error in Wasserstein $1$-distance between the true distribution and the numerical distribution of each algorithm. Calculating Wasserstein errors in dimensions more than one is a highly non-trivial task, but in our case, such a calculation becomes possible by exploiting the independence structure of our probability distribution (more details in the Appendix \ref{app:wasserstein}). We plot in Figure \ref{fig:Wasser_errors}  the Wasserstein error in the marginals as well as the total Wasserstein error. We have used 10 copies of the true samples to estimate the Monte Carlo error in the estimator of the $W^1$-error, the average error between the true samples is given as the black line while the grey area denotes one standard deviation around the average error. As we can see, consistent with the histograms, all the three algorithms are biased, with MLAm having the smallest bias. In addition, all the three algorithms suffer from slow convergence due to the ill-conditioning of $\Sigma$.

In order to increase the speed of convergence of the MLAm we now consider the barrier function
\begin{equation*} 
\psi_{2}(x) = \frac{1}{2}\psi_{1}(x)+\frac{1}{4}x^{\top}\Sigma^{-1}x,
\end{equation*}  
which incorporates the idea of preconditioning, and hence denoted as P-MLAm. Furthermore, to eliminate the bias we consider the Metropolis adjusted algorithm MAMLA. As discussed previously, this is an adjusted discretization of \eqref{eq:mirror1} and hence we use a slightly different barrier function given by  
\begin{equation*} 
\psi_{3}(x) = \frac{1}{2}\psi_{1}(x)+\frac{1}{4}x^{\top}\Sigma^{-1/2}x. 
\end{equation*} 
For both of the algorithms, there is no visible bias in the histograms, however a closer look of the Wasserstein error shows a small bias for P-MLAm. 




\begin{figure}[ht]
    \centering
    \begin{subfigure}[t]{0.32\linewidth}
        \includegraphics[width=\linewidth]{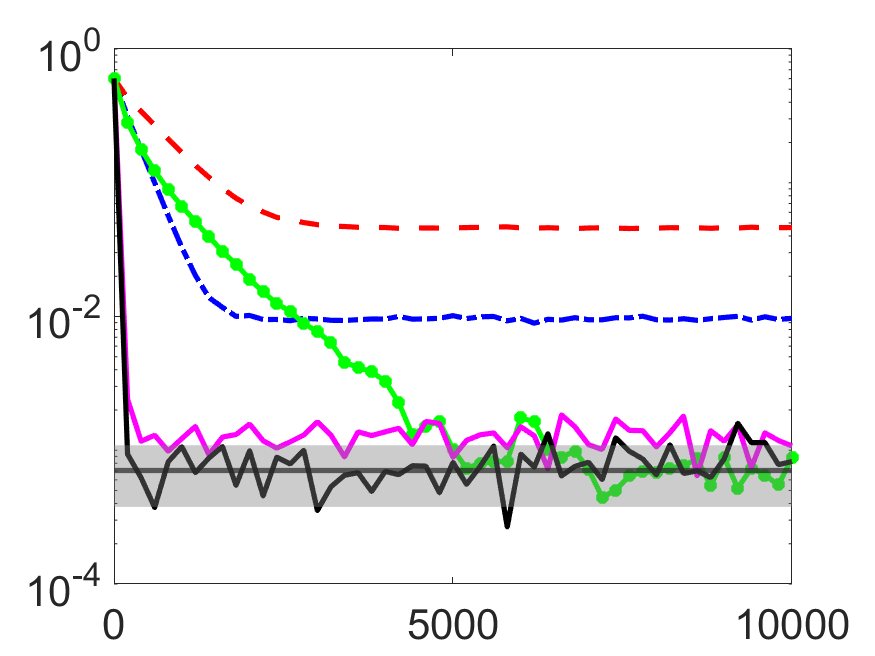}
        \caption{$W^{1}$ error for $x_{1}$ marginal }
    \end{subfigure}
    \begin{subfigure}[t]{0.32\linewidth}
        \includegraphics[width=\linewidth]{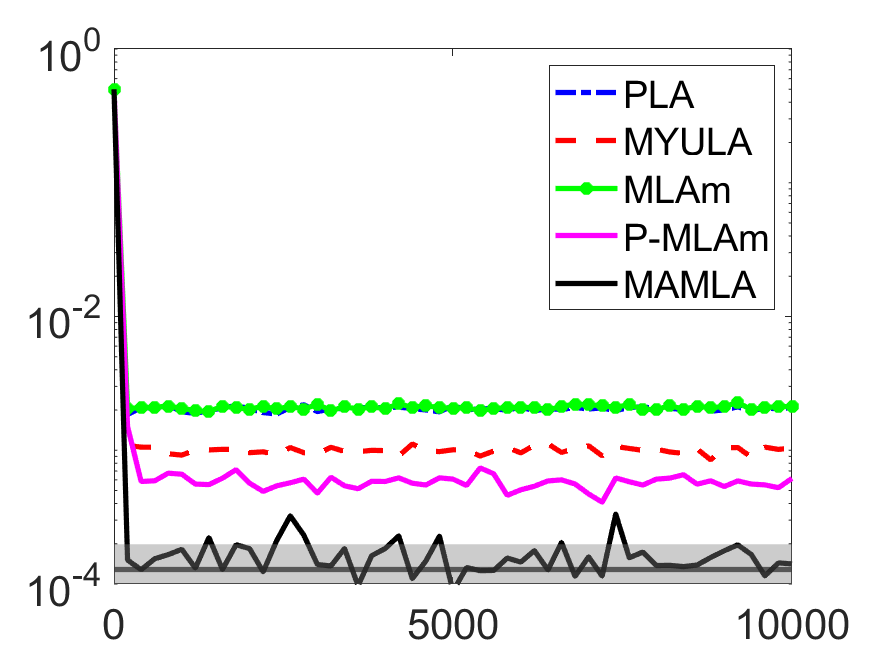}
         \caption{$W^{1}$ error for $x_{2}$ marginal}
    \end{subfigure}
    \begin{subfigure}[t]{0.32\linewidth}
        \includegraphics[width=\linewidth]{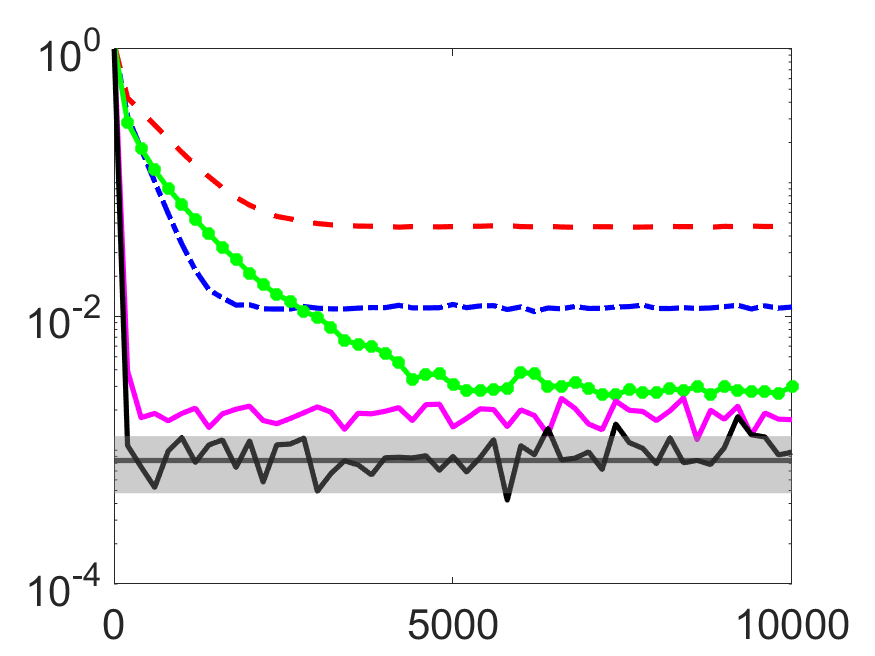}
         \caption{Total $W^{1}$ error}
    \end{subfigure}
    \caption{Errors for the different algorithms}
    \label{fig:Wasser_errors}
\end{figure}


\section{PDMPs based samplers: Unconstrained case}\label{sec3}
In this section, we  review some standard approaches for sampling from a density of the form \eqref{eq:prob} using PDMPs in the case where $\mathcal{M}=\mathbb{R}^{d}$.
\subsection{Introduction to piecewise deterministic Markov processes}
A PDMP is a continuous time Markov process $\{ {z}_t \}_{t \geq 0}$ whose trajectories follow  between jump times the deterministic dynamics given by the ODE  
\begin{equation}\label{dd} 
\frac{dz_t}{dt} = G(z_t), \quad z_0 = z.
\end{equation}
The jumps occur at random times $(T_i)_{i \geq 1}$ that have rate $\lambda(z_t)$. That is, the inter-jump times $\tau_i=T_i-T_{i-1}$ have law given by
\begin{equation}\label{tau-dist} 
\mathbb{P}(\tau_i \geq t| z_{T_{i-1}}=z) = \exp \left( -\int_{0}^t \lambda(\phi_s(z)) ds  \right).
\end{equation}
Here $\phi_s(z)$ denotes solution of \eqref{dd} at time $s$ with initial condition $z$. At jump times the process jumps from a state $z$ to $z'$ according to a transition kernel $Q(z'|z)$. A PDMP $\{ z_t \}_{t\geq0}$ is therefore characterized by the triplet $(G, \lambda, Q)$. 

The framework described above is very general and encompasses a number of important processes. Similarly to the case of SDE-based samplers, there are a lot of different choices for the triplet $(G, \lambda, Q)$ that would lead to dynamics that are ergodic with respect to \eqref{eq:prob}. We discuss in Section \ref{subsec:ex} some standard choices for which \eqref{dd} admits an explicit solution.

\subsection{Implementation}\label{subsec:implementation}
In order to be able to exactly simulate PDMPs, one needs: firstly  to solve the deterministic dynamics \eqref{dd} explicitly; secondly to exactly simulate the jump times with law \eqref{tau-dist}. It is indeed the case for PDMPs that arise in MCMC, such as the ZZS and the BPS, that the dynamics \eqref{dd} often admit an explicit solution. However, for a general $U(x)$, exact simulation of the inter-jump times is non-trivial. 

An approach to do this is Poisson thinning (see \cite{lewis1979simulation}). The idea behind Poisson thinning is that one can simulate an inhomogeneous exponential random variable with rate $m(s)$ as follows: Suppose that there exists a function $M$ such that $m(s) \leq M(s)$ for all $s\geq 0$. Define 
\begin{equation*}\label{tau-thinning}
\tau^* = \inf\left\{ t>0: \ \int_{0}^t M(s) \ ds   \geq \mathcal{E}\right\}, 
\end{equation*}
where $\mathcal{E} \sim \text{Exp}(1)$, and the proposed time $\tau^*$ is accepted with probability $m(\tau^*)/ M(\tau^*)$. In the context of PDMPs one can use Poisson 
thinning with $m(s) = \lambda(\phi_s(z))$ and 
selecting a function $M$ such that simulating 
$\tau^*$ is computationally efficient.  

We summarise  in Algorithm \ref{alg:PDMP-Sampler} how one would implement a general PDMP sampler assuming that an appropriate bound for $m(s)$ exists.
\begin{algorithm}[ht]
	\caption{General PDMP Sampler}
	\label{alg:PDMP-Sampler}
	\begin{algorithmic}[1]
		\REQUIRE $ z_0$ Initial state, $N$ number of skeleton points 
		\ENSURE $(T_k,z_{T_k})_{k=0}^N$, the skeleton
		\STATE Set $k=0$, $(T_0, z_{T_0}) = (0, z_0), \tau=0, z^*=z_0$,
		\WHILE{$k<N$}
            \STATE Let $M(s) = M(s;\tau,z_{T_k})$ be the computational bound of $m(s) = \lambda(\phi_{s+\tau}(z_{T_k}))$ 
            \STATE Draw $\mathcal{E} \sim \mathrm{Exp}(1)$
		\STATE Draw a candidate inter-jump time $\tau^*$ by solving 
		$$\tau^* = \inf\left\{t>0:\int_{0}^{t} M(s) \ ds \geq \mathcal{E}  \right\}$$
		\STATE Update $z^* = \phi_{\tau^*}(z^*)$, $\tau=\tau+\tau^*$
		\STATE Draw $u \sim \mathcal{U}(0,1)$ 
		\IF{ $u < \lambda(z^*)/M(\tau)$} 
                 \STATE $z^* \sim Q ({z}^*, \cdot)$,
                 $(T_{k+1}, {z}_{T_{k+1}}) = (T_k+\tau, {z}^*)$, \\
                 \STATE $\tau=0$, $k=k+1$
            \ENDIF

		\ENDWHILE 
	\end{algorithmic}
\end{algorithm}
Note that PDMPs are continuous time processes and the output from Algorithm~\ref{alg:PDMP-Sampler} is the skeleton points needed to recover the entire path of the process. One can obtain $N$ samples from this path by evaluating the path at equally spaced time points $t_k=kT/N$, where $T$ is the final time the PDMP has been simulated up until. These points can be calculated  using that
\begin{equation*}
    z_{t_k} = \phi_{t_k-T_{n(t_k)}}(z_{T_{n(t_k)}})
\end{equation*}
where $T_{n(t_k)}$ is the largest jump time before $t_k$.

\subsection{Examples of PDMPs} \label{subsec:ex}
We discuss two examples of PDMPs for which  the deterministic  dynamics \eqref{dd} admit explicit solutions. 

\subsubsection{Zig-Zag sampler}
For the ZZS, the process $z_t$ is written as $(x_t,v_t)$ where $x_{t}$ denotes position and $v_t$ the velocity. The velocity component $v_t$ takes values in $\{-1,1\}^d$ while $x_t$ takes values in $\mathbb{R}^d$. The ODE dynamics have constant velocity and are given by
\begin{align} \label{eq:zz_dyn}
    \frac{d}{dt}x_t = v_t, \quad \frac{d}{dt}v_t=0.
\end{align}
The jump rate is given by
\begin{equation} \label{eq:zz_jr}
    \lambda(x,v) = \sum_{i=1}^d \lambda_i(x,v), \quad \lambda_i(x,v) = (v_i\partial_iU(x))^+.
\end{equation}
Here $(\cdot)^+$ denotes the maximum with zero. 
The transition kernel $Q$ corresponds to coordinate reflections and is given by
\begin{align*}
    Q((dx',dv')|(x,v)) &= \sum_{i=1}^d \frac{\lambda_i(x,v)}{\lambda(x,v)} Q_i((dx',dv')|(x,v)), \\
     Q_i((dx',dv')|(x,v)) & = \delta_{(x,F_i(v))}(dx',dv'), \\
    F_i(v) &= (v_1,\ldots, -v_i, \ldots, v_d).
\end{align*}
To summarise, the dynamics follow a simple linear trajectory $\phi_t(x,v) = (x+tv,v)$  between jump times. At jump times one component of the velocity changes sign.
Under suitable conditions this process has a unique stationary distribution given by   $\rho = \pi \otimes \mathrm{Unif}(\{\pm 1\}^d) $ (see \cite{bierkens2019ergodicity}). 

The implementation of the skeleton of ZZS is given in Algorithm~\ref{alg:ZZS}. Note that a  simple bound for $\lambda$ when $U$ is $L$-gradient Lipschitz is given by 
\begin{equation*}
    M=\sum_{i=1}^d M_i, \quad M_i(s) = \lambda_i({x},{v}) + L \lVert {v}\rVert s.
\end{equation*}

\begin{algorithm}[ht]
	\caption{Zig-Zag Sampler (ZZS)}
	\label{alg:ZZS}
	\begin{algorithmic}[1]
		\REQUIRE $ ({x}_0,{v}_0$) Initial state, $N$ number of skeleton points, computational bounds $M_i$ 
		\ENSURE $(T_k,{x}_{T_k},{v}_{T_k})_{k=0}^N$, the skeleton
		\STATE Set $k=0$, $(T_0, x_{T_0},{v}_{T_0}) = (0, {x}_0,{v}_0), \tau=0, x^*=x_{T_0}$
		\WHILE{$k<N$}
        \STATE Let $M_i(s) = M_i(s;\tau,x_{T_k},v_{T_k})$ be a computational bound of $$m_i(s) = \lambda_i(x_{T_k}+(s+\tau) v_{T_k},v_{T_k})$$ 
        \STATE Draw $\mathcal{E}_i \sim \mathrm{Exp}(1)$ independently for $i=1,\dots,d$
        \STATE Draw $\tau_1^*,\ldots,\tau_d^*$ by solving $$\tau_i^* = \inf\left\{t>0:\int_{0}^{t} M_i(s) \ ds \geq \mathcal{E}_i  \right\}$$
        \STATE Set $i_0=\arg\min_{i=1,\ldots,d} \tau_i^*$ and $\tau^*=\tau_{i_0}^*$.
        \STATE $(\tau,x^*) = (\tau+\tau^*,x^*+\tau^* {v}_{T_k})$.
        \STATE Draw $u \sim \mathcal{U}(0,1)$
        \IF{ $ u < \frac{m_{i_0}(\tau^*)}{M_{i_0}(\tau^*)} $}
        \STATE Set
        \begin{align*}
        {v}_{T_{k+1}} &= F_{i_0}({v}_{T_k}) , \\
        {x}_{T_{k+1}}&= x^*, \\
        T_{k+1}&=T_k+\tau,  \\
        k&=k+1.
        \end{align*}
        \ENDIF
		\ENDWHILE 
	\end{algorithmic}
\end{algorithm}
\subsubsection{Bouncy particle sampler}
For the BPS, the process $z_{t}$ is written as $(x_{t},v_{t})$ where  $x_{t}$ denotes position and $v_{t}$ velocity. Now both the position and velocity components take values in $\mathbb{R}^{d}$. The ODE dynamics again follow \eqref{eq:zz_dyn} while the jump rate is given by 
\begin{equation} \label{eq:bps_jr}
    \lambda(x,v)  = (v^\top\nabla U(x))^++\gamma(x),
\end{equation}
where $\gamma(x)>0$ is a positive function (typically taken to be constant and referred to as the refreshment rate). 
The transition kernel $Q$, which corresponds to contour reflections and refreshments, is given by
\begin{align*}
    Q\left((dx',dv')|(x,v)\right) &
    = \frac{(v^\top\nabla U(x))^+}{\lambda(x,v)}\delta_{(x,R(x)v)}(dx',dv')+\frac{\gamma(x)}{\lambda(x,v)}\delta_{x}(dx') \nu(dv') \\
    R(x)v &= v-2\frac{v^\top \nabla U(x)}{\lVert \nabla U(x)\rVert^2}\nabla U(x).
\end{align*}
Here $\nu$ is the velocity marginal of the stationary distribution and is usually taken either to be the normal distribution on $\mathbb{R}^d$ or the uniform distribution on the unit sphere of $\mathbb{R}^d$. To summarise, the dynamics, similar to the case of the ZZS, follow a simple linear trajectory $\phi_t(x,v)=(x+tv,v)$ between jump times. At a jump the velocity is either replaced by a new sample from $\nu$ or is reflected according to the contours of $\nabla U$. Under suitable conditions, this process has a unique stationary distribution given by $\rho(dx,dv)= \pi(dx) \otimes \nu(dv)$ (see \cite{durmus20,deligiannidis2019exponential}).

For implementation, a simple bound for $\lambda$ when $\gamma(x)=\gamma$ is constant and $U$ is $L$-gradient Lipschitz is given by 
\begin{equation*}
M(s) =\lambda({x},{v}) + L \lVert {v}\rVert^{2} s. 
\end{equation*}

\subsection{Subsampling}\label{sec:subsampling}

One of the main advantages of PDMPs over SDE-based approaches is that PDMPs allow for exact subsampling \citep{bierkens2019zig} which means that the process can simulate exact samples from $\pi$ while using an unbiased estimator of the gradient $\nabla U$.  
More precisely, in the context of Bayesian statistics $\pi(x)=p(x|y)$ is the Bayesian posterior density of $x$ given some data $y$. We define $\pi_0$ to be the density of the prior, $f(y|x)$ as the likelihood, and $U^j(x) =-\log\pi_0(x)-K\log f(y^j|x)$, with $y=(y^1,\ldots, y^K)$ being the data.
Then the Bayesian posterior can be written in the form \eqref{eq:prob} with 
\begin{equation}\label{eq:bayesian_setting}
    U(x) = \frac{1}{K}\sum_{j=1}^K U^j(x).
\end{equation}
In general, there are many cases of practical interest \citep{bierkens2019zig,WT11} where the function $U$ in \eqref{eq:prob} is of the form
\begin{equation} \label{eq:sum_form}
    \partial_i U({x})  = \frac{1}{K}\sum_{j=1}^K E_i^j({x}), \quad i=1,\ldots, d
\end{equation} 
for some functions $E_i^j$, where $K$ can be very large.  As we have seen in Section \ref{subsec:ex} implementing the ZZS (or BPS)  requires evaluating $\nabla U$ at every step of the algorithm.  In the setting \eqref{eq:bayesian_setting}, evaluating $\nabla U$ is prohibitively expensive as it requires a full pass over the data. 
We can avoid this by considering the Zig-Zag sampler with subsampling (ZZ-SS), which is a version of the ZZS with suitably chosen jump rate $\lambda$. This sampler can be implemented without requiring a full pass through the data at each step.

The ZZ-SS is a PDMP with the same deterministic dynamics and transition kernel as the ZZS but with jump rate given by
\begin{equation*}
    \lambda (x,v) = \sum_{i=1}^d \lambda_i(x,v),\quad  \lambda_i(x,v) = \frac{1}{K}\sum_{j=1}^K \lambda_i^j(x,v), \quad \lambda_i^j(x,v) = \left(v_iE_i^j(x)\right)^+.
\end{equation*}
Note that this switching rate $\lambda$ is larger than the canonical rate used by the ZZS. This means that the process will have more jumps per unit time, and therefore it will require simulating more jumps to reach the same level of integration time compared to ZZS. Hence it is not directly obvious why this formulation is advantageous since if naively implemented it still requires a full pass through the data at each step of the algorithm, and it also leads to poorer mixing. However,  it is now very easy to construct an unbiased estimator of $\lambda_{i}$ by considering
\[
\widehat{\lambda}_{i}(x,v)=\left( v_{i}E^{J}_{i}(x) \right)^{+}, \quad J \sim \mathrm{Unif}(\{1,\ldots,K\})
\]
which only needs evaluation of a single $E_{i}^{j}(x).$ This combined with the properties of Poisson thinning, allows one to implement ZZ-SS without requiring a full pass of the data at each step. In practice, this corresponds to Algorithm \ref{alg:ZZ-SS} which only differs from Algorithm \ref{alg:ZZS} by replacing $\lambda$ with its unbiased estimator in the acceptance rate of the Poisson thinning. 

In principle, when constructing the unbiased estimator one can set $E_i^j=\partial_iU^j$, however as shown in \cite{bierkens2019zig} it is more efficient to use the idea of control variates. This relies on finding a suitable reference point $\hat{x}$ and then defining
\begin{equation}\label{eq:Eij_definition}
    E_i^j(x) = \partial_i U(\hat{x}) +\partial_iU^j(x)-\partial_iU^j(\hat{x}).
\end{equation}
The reference point recommended in \cite{bierkens2019zig} is the mode of the distribution, this can be found using a numerical optimisation routine. Although \eqref{eq:Eij_definition} involves the full derivative $\partial_i U$, this is only evaluated at the reference point $\hat{x}$ so can be calculated once prior to running the algorithm. A simple bound for $\lambda_i^j$ when $U$ is $L_p$-gradient Lipschitz with respect to the $\lVert \cdot \rVert_p$ norm is given by
\begin{equation}\label{eq:ZZ-SS bound}
\widehat{M}_i(s) = (v_i\partial_iU(\hat{x}))^+ +L_p \lVert x-\hat{x}\rVert_p + L_p \lVert v\rVert_p  s.
\end{equation}

\begin{algorithm}[ht]
	\caption{Zig-Zag Sampler with subsampling (ZZ-SS)}
	\label{alg:ZZ-SS}
	\begin{algorithmic}[1]
		\REQUIRE $ {x}_0,{v}_0$ Initial state, $N$ number of skeleton points, computational bounds $M_i$ 
		\ENSURE $(T_k,{x}_{T_k},{v}_{T_k})_{k=0}^N$, the skeleton
		\STATE Set $k=0$, $(T_0, x_{T_0},{v}_{T_0}) = (0, {x}_0,{v}_0), \tau=0, x^*=x_{T_0}$
		\WHILE{$k<N$}
        \STATE Let $M_i(s) = M_i(s;\tau,x_{T_k},v_{T_k})$ be a computational bound of $$m_i^j(s) = \lambda_i^j(x_{T_k}+(s+\tau) v_{T_k},v_{T_k})$$ 
        \STATE Draw $\mathcal{E}_i \sim \mathrm{Exp}(1)$ independently for $i=1,\dots,d$
         \STATE Draw $\tau_1^*,\ldots,\tau_d^*$ by solving $$\tau_i^* = \inf\left\{t>0:\int_{0}^{t} M_i(s) \ ds \geq \mathcal{E}_i  \right\}$$
        \STATE Set $i_0=\arg\min_{i=1,\ldots,d} \tau_i^*$ and $\tau^*=\tau_{i_0}^*$.
        \STATE $(\tau,x^*) = (\tau+\tau^*,x^*+\tau^* {v}_{T_k})$.
        \STATE Draw $J\sim \mathrm{Unif}(\{1,\ldots,K\})$.
        \STATE Draw $u \sim \mathcal{U}(0,1)$
        \IF {$u < \frac{m_{i_0}^J(\tau^*)}{M_{i_0}(\tau^*)}$}
        \STATE Set
        \begin{align*}
        {v}_{T_{k+1}} &= F_{i_0}({v}_{T_k}) , \\
        {x}_{T_{k+1}}&= x^*, \\
        T_{k+1}&=T_k+\tau,  \\
        k&=k+1.
        \end{align*}
        \ENDIF
		\ENDWHILE 
	\end{algorithmic}
\end{algorithm}


\section{Mirror PDMPs}\label{mirror-PDMPs}
We now turn our interest to sampling \eqref{eq:prob} in the case where $\mathcal{M}$ is a proper convex subset of $\mathbb{R}^{d}$ using PDMPs. In principle, a starting point to construct a  PDMP $z_{t}$ that is ergodic with respect to \eqref{eq:prob}  is to introduce some dynamics \eqref{dd} that would ensure that the relevant part of the process (for example $x_{t}$ in ZZS and BPS) remains in $\mathcal{M}$. One would then need to modify the jump rates $\lambda$ and jump transition kernel $Q$ so that the process is ergodic with respect to \eqref{eq:prob}.  

As we saw in Section \ref{sec:num_ill} the mirror map better exploits the geometrical structure of $\mathcal{M}$, and for this reason we will combine ideas of PDMPs and mirror maps. In order to achieve this, we  consider  ${\zeta}_t=\nabla\psi({z}_t)$ where $\psi$ is a barrier function. Now because of the properties of $\psi$, it is sufficient to ensure that $\zeta_{t}$ is invariant with respect to the push-forward distribution $\tilde{\pi}$ defined through \eqref{pushforward-potential}. Note that this is an unconstrained distribution and hence one can choose any PDMP which is ergodic with respect to $\tilde{\pi}$ to define $\zeta_{t}$. One can then obtain samples which are ergodic with respect to \eqref{eq:prob} by applying $\nabla \psi^{*}$ to samples of $\zeta$. We  now discuss mirror ZZS which is an implementation based on the ZZS, similarly one can derive the mirror BPS.

\subsection{The Mirror Zig-Zag Sampler}  

As discussed above we start by considering a ZZS defined in the dual space, in particular we have $({\zeta}_t,{v}_t)$ with ODE dynamics
\begin{align*}
    \frac{d}{dt}{\zeta}_t = {v}_t, \quad \frac{d}{dt}{v}_t=0.
\end{align*}
The rate is given by
\begin{equation*}
\tilde{\lambda}({\zeta},{v}) = \sum_{i=1}^d \tilde{\lambda}_i({\zeta},{v}), \quad \tilde{\lambda}_i({\zeta},{v}) = (v_i\partial_iV({\zeta}))^+.
\end{equation*}
Recall $V$ is given by \eqref{pushforward-potential}. Finally, the transition kernel corresponds to coordinate reflections. Transforming this PDMP to the primal space gives rise to the Mirror Zig-Zag Sampler (MZZS), more precisely we set $ ({x}_t, {v}_t)  =  ( \nabla \psi^*({\zeta}_t), {v}_t )$. The procedure to generate the skeleton of $({x}_t,{v}_t)$ is given in Algorithm \ref{alg:MZZS}. To generate the samples, we first create the skeleton $(T_k,{\zeta}_{T_k},{v}_{T_k}) = (T_k,\nabla\psi({x}_{T_k}),{v}_{T_k})$ in the dual space. By linearly interpolating between the points ${\zeta}_{T_k}$, we can reconstruct the entire continuous path, hence we can obtain samples ${X}_{t_n} = \nabla\psi^*({\zeta}_{t_n})$ where $t_n$ are equally spaced time points between $0$ and the final time. 

Similarly to the case of the ZZS, if $V$ is $L_V$-gradient Lipschitz then we can use the bound
\begin{equation}\label{eq:MZZS_bound}
    \tilde{M}=\sum_{i=1}^d \tilde{M}_i, \quad \tilde{M}_i(s) = \tilde{\lambda}_i({x},{v}) + L_V \lVert {v}\rVert s
\end{equation}
to perform Poisson thinning. Note that, in principle, even though one could derive a general expression for $L_V$ in terms of bounds on the derivatives of $U$ and $\psi^*$, this expression would be very cumbersome and typically a case by case approach will provide a more useful bound. For all the experiments in this paper we will apply a case by case approach.

\begin{remark}\label{rmk:mpdmp_generator}
    Similarly to the discussion on mirror-based SDEs it is possible to transform the MZZS into a processes on the primal space. This is itself a PDMP with the following deterministic dynamics
  \begin{equation*}\label{MZZP-dynamic}
       \begin{cases}
           \dot{{x}}_t = \nabla^2 \psi({x}_{t})^{-1} {v_t}, \\
           \dot{{v}}_t = 0.
       \end{cases}
   \end{equation*}
   In addition the rate $\lambda$ satisfies \begin{equation*}\label{MZZP-rates}
       \lambda({x}, {v}) = \widetilde{\lambda}(\nabla\psi({x}),{v})
   \end{equation*} 
   while the transition kernel remains unchanged.

     Observe that these deterministic dynamics can be viewed as a gradient flow in the direction $v$ on the Riemmanian manifold with metric $\nabla^2\psi$ which is described in Remark~\ref{rem:remannian_manifold}. Therefore, one can view MZZS as the Riemannian analogue of ZZS where the constant velocity dynamics have become gradient flows with constant direction.
\end{remark}

\begin{remark}
    Note that analogously to Section \ref{sec:subsampling} one can suitably modify the MZZS to the mirror Zig-Zag with subsampling (MZZ-SS). 
\end{remark}

\begin{algorithm}[ht]
	\caption{Mirror Zig-Zag Sampler}
	\label{alg:MZZS}
	\begin{algorithmic}[1]
		\REQUIRE $ ({x}_0,{v}_0)$ Initial state, $N$ number of skeleton points 
		\ENSURE $(T_k,{x}_{T_k},{v}_{T_k})_{k=0}^N$, the skeleton
		\STATE Set $k=0$, $(T_0, {\zeta}_{T_0},{v}_{T_0}) = (0, \nabla\psi({x}_0),{v}_0), \tau=0, {\zeta}^*={\zeta}_{T_0}$,
		\WHILE{$k<N$}
        \STATE Let $\widetilde{M}_i(s) = \widetilde{M}_i(s;\tau,\zeta_{T_k},v_{T_k})$ be a computational bound of $$\widetilde{m}_i(s) = \widetilde{\lambda}_i(\zeta_{T_k}+(s+\tau) v_{T_k},v_{T_k})$$ 
        \STATE Draw $\mathcal{E}_i \sim \mathrm{Exp}(1)$ independently for $i=1,\dots,d$ 
         \STATE Draw $\tau_1^*,\ldots,\tau_d^*$ by solving $$\tau_i^* = \inf\left\{t>0:\int_{0}^{t} \widetilde{M}_i(s) \ ds \geq \mathcal{E}_i  \right\}$$
        \STATE Set $i_0=\arg\min_{i=1,\ldots,d} \tau_i^*$ and $\tau^*=\tau_{i_0}^*$.
        \STATE $(\tau,\zeta^*) = (\tau+\tau^*,\zeta^*+\tau^* {v}_{T_k})$.
        \STATE Draw $u \sim \mathcal{U}(0,1)$ 
        \IF {$u < \widetilde{m}_{i_0}(\tau)/\widetilde{M}_{i_0}(\tau)$}
        \STATE Set 
        \begin{align*}
        {v}_{T_{k+1}} &= F_{i_0}({v}_{T_k}) , \\
        {x}_{T_{k+1}}&= \nabla\psi^*({\zeta}^*), \\
        T_{k+1}&=T_k+\tau, \\
        k & =k+1
        \end{align*}
        \ENDIF
     \ENDWHILE 
	\end{algorithmic}
\end{algorithm}

\begin{remark}
  The proposed algorithm MZZS is similar in spirit to MLAa, see \eqref{eq:MLD}, in the sense that both algorithms operate in the dual space using the transformed potential V given by \eqref{pushforward-potential} and map back to the primal space using $\nabla \psi^*$. In principle, one could construct a mirror PDMP similarly to MLAm by incorporating the geometry of the Riemannian manifold in the dual space. However, doing this is not trivial and the corresponding equations would not typically admit closed form solutions.
\end{remark}

\begin{remark}
    One can construct a mirror Bouncy Particle Sampler (MBPS) analogously to MZZS. If $V$ is $L_V$-gradient Lipschitz and the refreshment rate $\gamma$ is constant then we can use the bound
    \begin{equation}\label{eq:MBPS_bound}
        \tilde{M}(s) =\tilde{\lambda}({x},{v}) + L_V \lVert {v}\rVert^{2} s. 
\end{equation}
\end{remark}

\subsubsection{Numerical illustration continued}\label{sec:num_ill_mzzs}
\begin{figure}[ht]
    \centering
    \begin{subfigure}[t]{0.32\linewidth}
        \includegraphics[width=\linewidth]{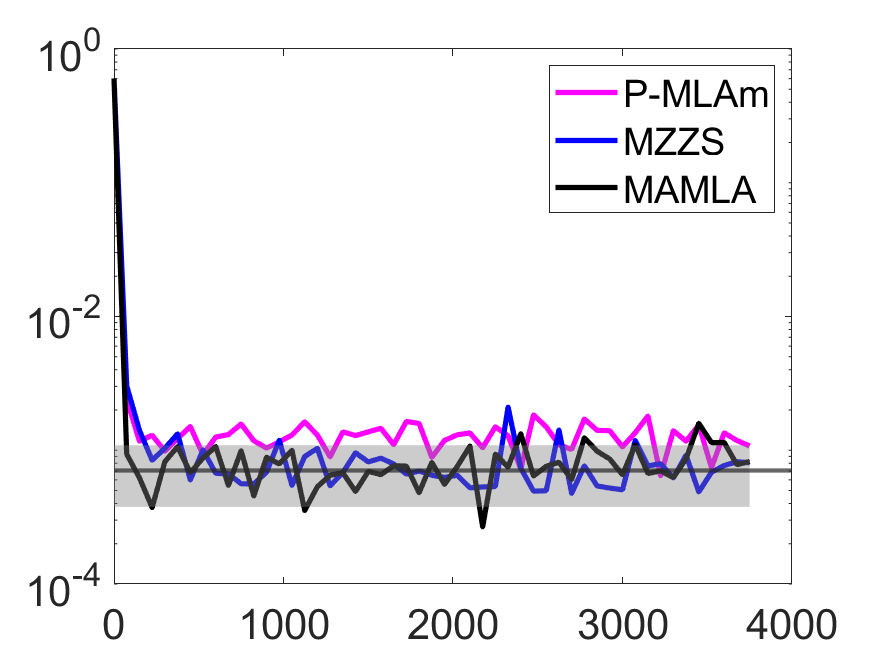}
        \caption{$W^{1}$ error for $x_{1}$ marginal }
    \end{subfigure}
    \begin{subfigure}[t]{0.32\linewidth}
        \includegraphics[width=\linewidth]{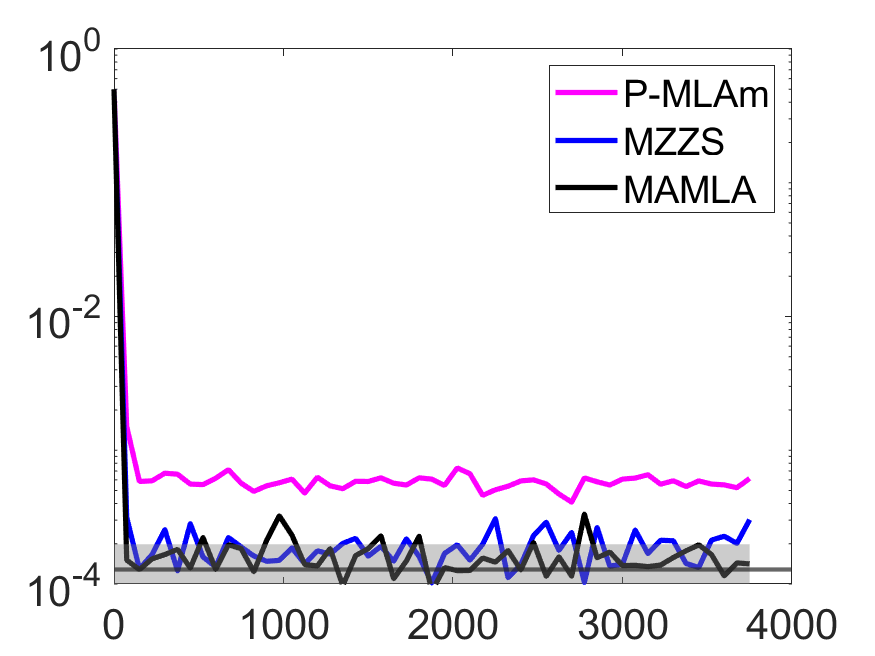}
        \caption{$W^{1}$ error for $x_{2}$ marginal } 
    \end{subfigure}
    \begin{subfigure}[t]{0.32\linewidth}
        \includegraphics[width=\linewidth]{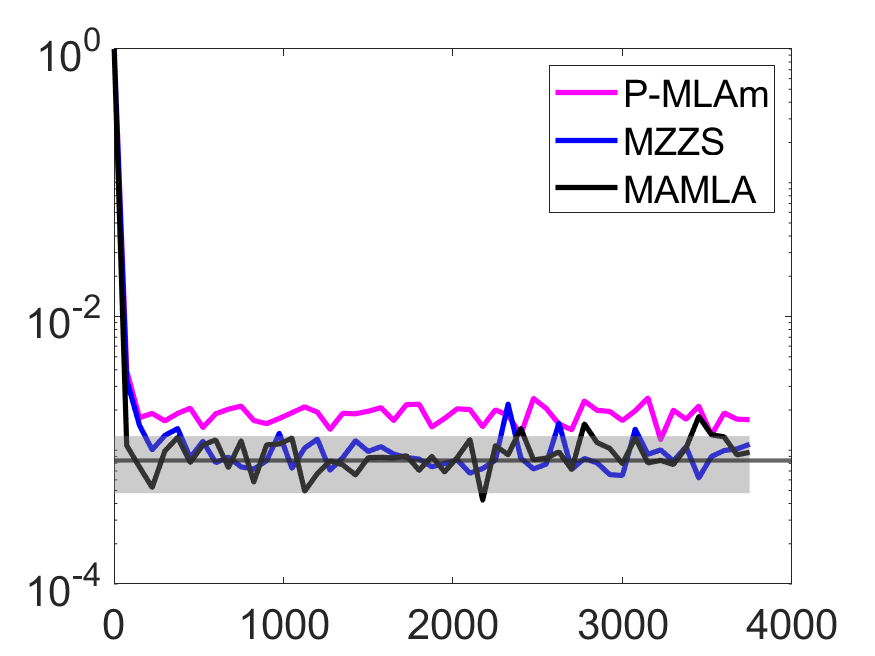}
         \caption{Total $W^{1}$ error}
    \end{subfigure}
    \caption{Errors for the different algorithms}
    \label{fig:Wasser_errorsZZS}
\end{figure}

We now return to the numerical illustration investigated in Section \ref{sec:num_ill} and include the MZZS with barrier function $\psi_{3}$. As we saw in Section \ref{sec:num_ill}, P-MLAm  exhibits the fast convergence with small  bias, while MAMLA converges with similar speed but with no bias.  Therefore, we only compare these mirror algorithms with MZZS. To make this comparison fair,  we rescale time so that the cost of simulating one unit of integration time for all those algorithms is roughly equal.  
We plot the $W^{1}$ error for these algorithms at fixed points of integration time in Figure \ref{fig:Wasser_errorsZZS} using $10^{6}$ independent copies. As we can see, all the algorithms converge with the same speed, but notably in the case of the MZZS and MAMLA, the methods are  unbiased as they achieve the same bias as the Monte Carlo error.

\subsection{Ergodicity of the Mirror Zig-Zag Sampler}
We now discuss conditions under which the MZZS is ergodic with respect to the probability distribution
\begin{equation*}
    \rho \propto \pi \otimes \mathrm{Unif}(\{\pm 1\}^d).
\end{equation*}
Here $\pi$ is given by \eqref{eq:prob}. More precisely, we discuss conditions under which
\begin{equation} \label{eq:erg}
        \lim_{t \to \infty} \| \mathbb{P}[({x}_t, {v}_t) \in A]- \rho(A) \|_{\mathrm{TV}} = 0,  
    \end{equation}
     for any deterministic initial conditions \  $({x}, {v}) \in \mathcal{M} \times \{-1,1\}^d$ and $A \in \mathcal{B}(\mathcal{M} \times \{-1,1\}^d)$. We will impose the following assumptions:

\begin{asmp}\label{ass:ergodicity}
    \begin{enumerate}[label=\textbf{A}\arabic*.]
        \item The function $V$ is three times continuously differentiable, i.e. $V\in \mathcal{C}^3(\mathbb{R}^d)$. \label{ass:regularity}
        \item The function $V$ has a non-degenerate local minimum, i.e. there exists a point $\zeta_0$ such that $\nabla V(\zeta_0)=0$ and $\nabla^2 V(\zeta_0) \succ 0$. \label{ass:nondegenerate}
        \item The function $V$ satisfies the growth condition
        \begin{equation*}
            V(\zeta) \geq c\log( \norm{\zeta}) -c'
        \end{equation*}
        for some $c>d$ and $c'\in\mathbb{R}$. \label{ass:growth}
    \end{enumerate}
\end{asmp}

Under these assumptions we directly employ the results of \cite[Theorem~1]{bierkens2019ergodicity} for the ZZS $(\zeta_t,v_t)$ to obtain the following theorem.
\begin{thm}
 If Assumption~\ref{ass:ergodicity} holds then $(\zeta_t,v_t)$ is invariant and ergodic with respect to $\tilde{\rho} \propto \tilde{\pi} \otimes \mathrm{Unif}(\{\pm 1\}^d)$, which in turn implies that  \eqref{eq:erg} holds for $(x_{t},v_{t})$.
\end{thm}

We now discuss each of these assumptions in turn.
\begin{enumerate}
    \item Assumption \ref{ass:regularity} can be ensured by requiring sufficient regularity of $U$ and $\psi$, indeed, if $U\in\mathcal{C}^3(\mathcal{M})$ and $\psi\in\mathcal{C}^5(\mathcal{M})$ then $V$ has the desired regularity. 
    \item Assumption \ref{ass:nondegenerate} is a technical condition required for the proof of ergodicity in \cite{bierkens2019ergodicity}.
    \item If we assume that $U$ satisfies
    \begin{equation}\label{eq:growth_U}
            U(x) \geq c_U\log( \norm{x}) -c_U'
        \end{equation}
        for some $c_U\geq 0$ and $c_U'\in\mathbb{R}$ then Assumption \ref{ass:growth} follows if
        \begin{equation}\label{eq:growth_psi}
            c_U\log(\norm{\nabla\psi^*(\zeta)})+\log\det(\nabla^2\psi(\nabla\psi^*(\zeta))) \geq c\log(\norm{\zeta})-c'
        \end{equation}
         for some $c>d$ and $c'\in\mathbb{R}$.
        Note that if $\mathcal{M}$ is a compact set then the growth condition \eqref{eq:growth_U} on $U$ follows from continuity of $U$ with $c_U=0$ and \eqref{eq:growth_psi} reduces to 
        \begin{equation*}
            \log\det(\nabla^2\psi(\nabla\psi^*(\zeta))) \geq c\log(\norm{\zeta})-c'.
        \end{equation*}
        On the other hand, if $\mathcal{M}$ is an unconstrained space then it is sufficient to have that $\psi$ is $m$-strongly convex, $U$ satisfies \eqref{eq:growth_U} with $c_U>d$ and $\nabla \psi^*(\zeta)$ grows at least linearly.
\end{enumerate}


\section{Numerics and discussions}\label{sec4}
We now present experimental results demonstrating the performance of the MZZS in a number of different examples. In addition, we compare its performance to state-of-the-art samplers. Throughout all of our experiments unless otherwise stated, we set the parameters for the algorithms as follows: For MLAa we use the step size $1/L_V$ with $L_V$ being the gradient-Lipschitz constant of $V$, for MLAm we use the step size $1/\tilde{L}$ where $\tilde{L}=L_V^S+ 2M_\psi L_V^L$ with $L_V^S$ being the relative smoothness constant of $U$, $M_\psi$ being the self-concordance constant of $\psi$ and $L_V^L$ being the relative Lipschitz constant of $U$ with respect to $\psi$(see \cite{ahn2021efficient} for details). 
We implement MZZS using the bound $\tilde{M}$ given by \eqref{eq:MZZS_bound}, and MBPS using the bound $\tilde{M}$ given by \eqref{eq:MBPS_bound}. For MBPS the refreshment rate has been manually tuned to obtain the best results. For MAMLA, we follow the approach used in the numerical illustration and run the algorithm adaptively, tuning the step size to achieve an acceptance rate of 25\%. This value is determined empirically based on the Gamma distribution example. Details on the derivation of the different constants will be given in the Appendix \ref{app:lips_const}.

\subsection{Gamma Distribution}\label{numeric2} 
We consider the Gamma distribution supported on $(0, +\infty)^d$ and defined by  
\begin{equation*}
\pi({x}) \propto \prod_{i=1}^{d}x_{i}^{\alpha_i-1} e^{-\beta_i x_{i}} = e^{-U({x})}, \quad U({x}) = \sum_{i=1}^{d}\beta_i x_{i} - (\alpha_i-1)\log(x_{i}).
\end{equation*}
The parameters satisfy $\alpha_i > 0$ and $\beta_i >0$ for all $i$. This distribution is useful in inverse problems, particularly as a prior when positivity is assumed for the quantity of interest. 

This distribution is particularly challenging to simulate using Langevin dynamics since the gradient of $U$ is non-Lipschitz. One could follow the approach of \cite{Savvas2023} by introducing a regularization of $U$ leading to a gradient Lipschitz approximation as well as a reflection to deal with the positivity constraint. However, the introduction of regularization introduces bias and we thus expect this approach not to be competitive. We will hence only compare MZZS  with various mirror Langevin for the barrier function $\psi$  defined in Table \ref{tab:barriers} for the positive orthant $(a_{i}=0, \ \forall i)$.

\begin{figure}[ht]
    \centering
    \begin{subfigure}[t]{0.49\linewidth}
        \includegraphics[width=\linewidth]{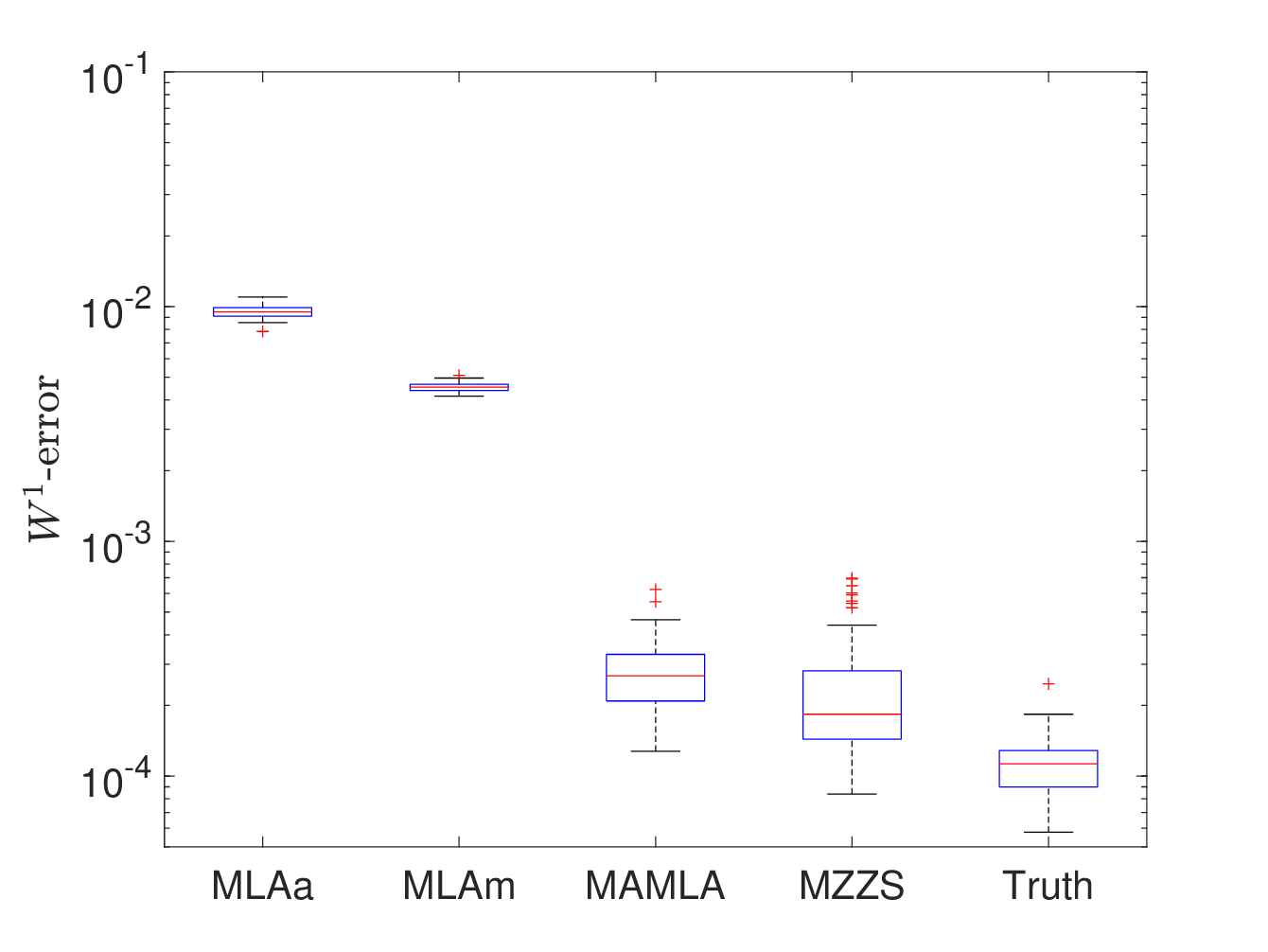} 
        \caption{$d=1$} 
    \end{subfigure}
    \begin{subfigure}[t]{0.49\linewidth}
        \includegraphics[width=\linewidth]{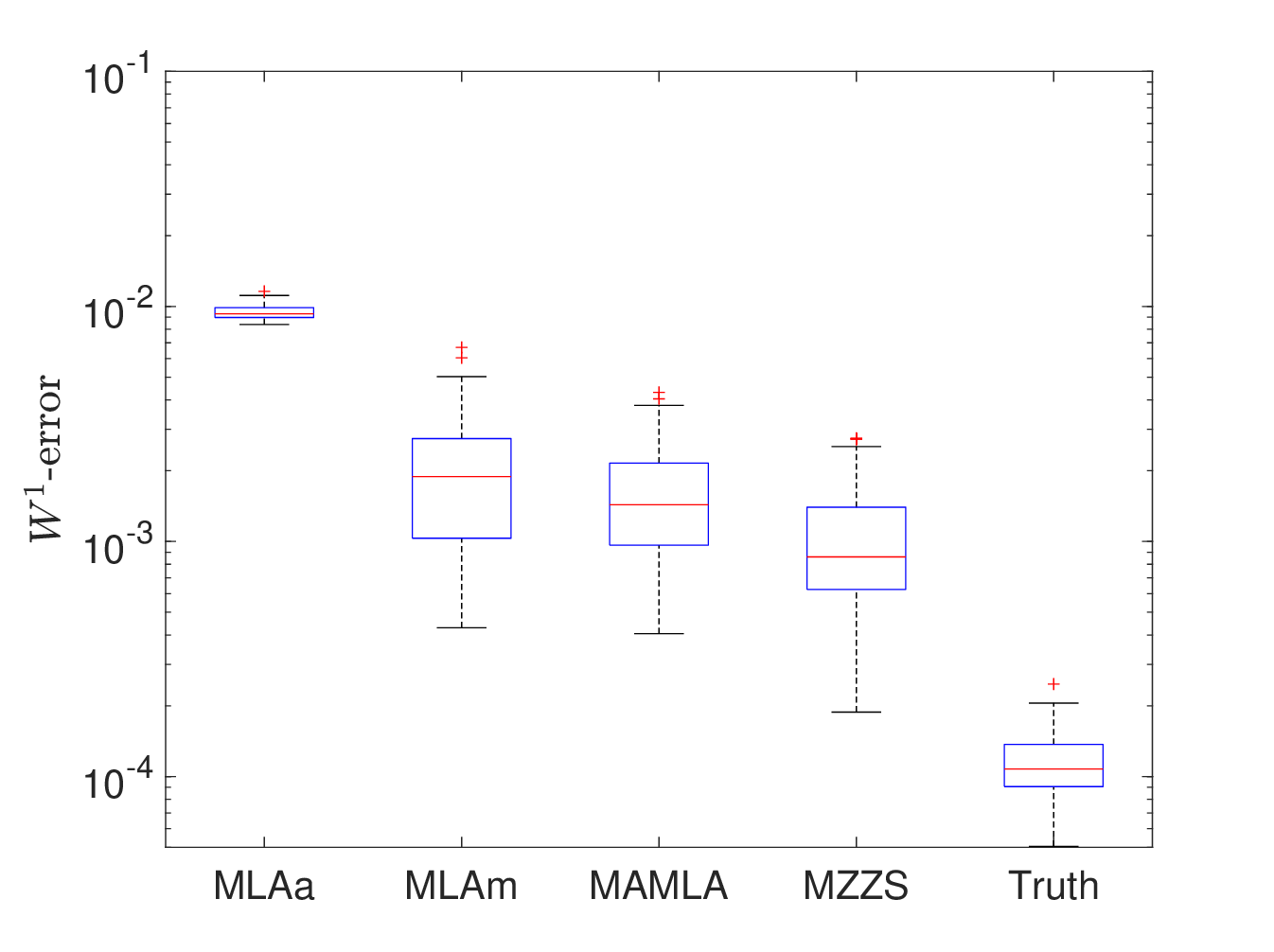} 
        \caption{$d=100$} 
    \end{subfigure}
    \caption{Box plots of $100$ realisations of the $W^1$ error of MLAa, MLAm, MAMLA and MZZS using $5\times 10^6$ gradient evaluations.}
    \label{fig:gamma_box_plots}
\end{figure}

For the simulations we consider dimensions $d=1$ and $d=100$ with shape parameters $\alpha_i=3$, and rate parameters $\beta_i=10$ for all $i$. We use $5\times 10^6$ gradient evaluations to generate $5\times 10^6$ samples from MLAa, MLAm, MAMLA and MZZS. For each set of samples we calculate the Wasserstein-1 error between the empirical distribution of the first component and the corresponding true samples of the Gamma distribution. This is repeated $100$ times and Figure~\ref{fig:gamma_box_plots} gives box plots of these Wasserstein errors.  We have also included a box plot showing the error between copies of the true samples which we use to measure the Monte Carlo error coming from the estimation of the Wasserstein error.
As we see from Figure~\ref{fig:gamma_box_plots} the MZZS leads to the smallest bias of the algorithms, due to the correlation between the samples there is more error than that observed from using independent true samples.

\subsection{Truncated Gaussian}\label{numeric1}
Our objective is to sample from the truncated Gaussian distribution in dimension $d = 10$ and defined on the open set ${\mathcal{M}} = (0,5) \times (0,0.5)^9$ by $\pi(x) \propto e^{-U(x)}$ with $U(x) = \displaystyle \frac{1}{2}x^\top \Sigma^{-1}x$. The covariance $\Sigma$ of the target is defined by $\Sigma_{ij} = (1+|i-j|)^{-1}.$ We use the mirror map $\psi$   defined in Table \ref{tab:barriers} for the rectangle.
We compare both the performance of MZZS and MBPS to MLAa, MLAm and MAMLA. Furthermore, for the special case of the truncated Gaussian distribution, Wall Hamiltonian Monte Carlo (WHMC), (\cite{pakman2014exact}) provides an exact way to sample and thus we consider the results obtained by it as the ground truth.

\begin{figure}[ht]
	\centering 
	\includegraphics[scale=0.3]{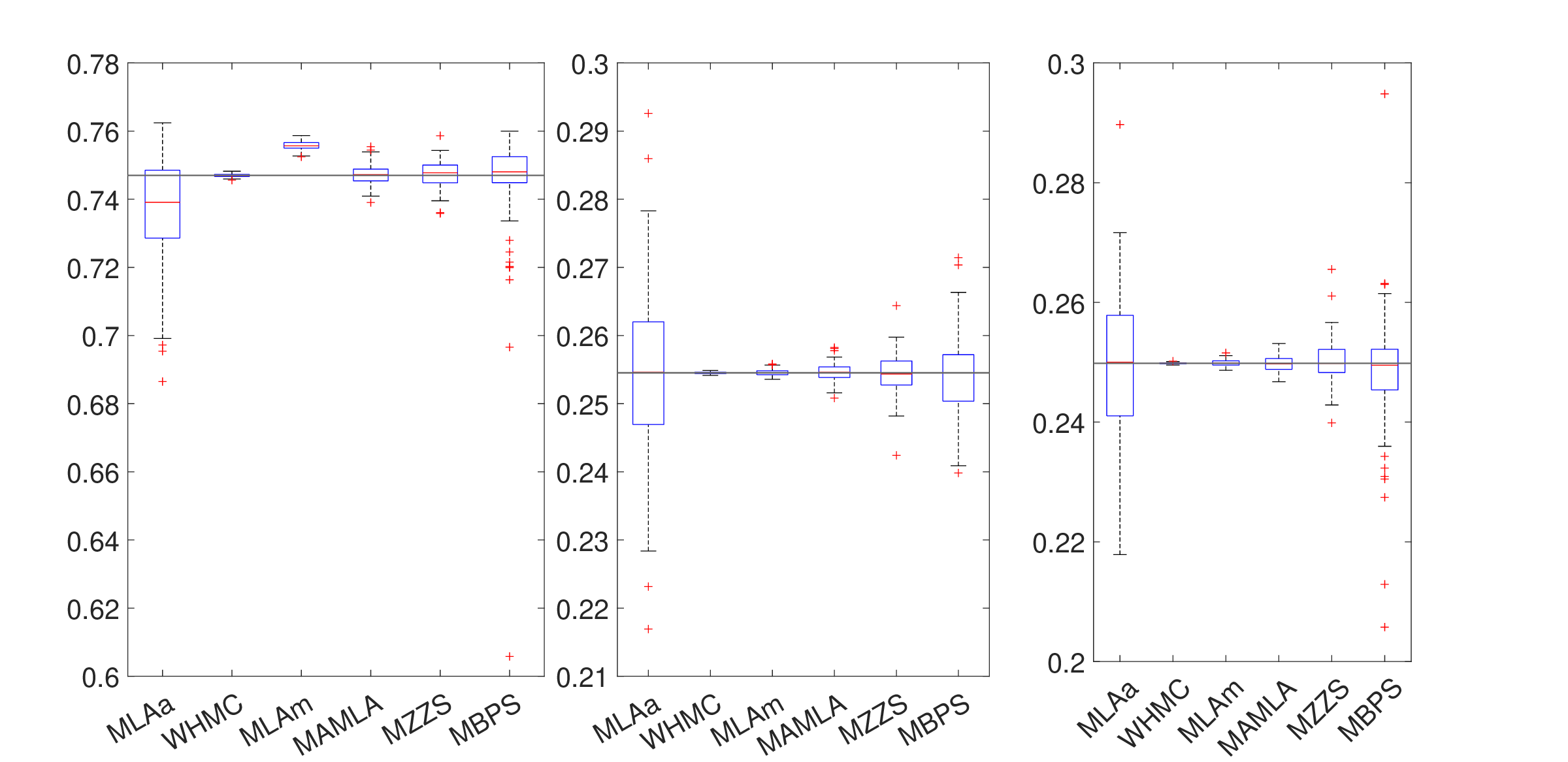} 
	\caption{Boxplots of the expectations $\mathbb{E}[x_1], \ \mathbb{E}[x_2]$ and $\mathbb{E}[x_3]$ for the truncated Gaussian distribution in 10 dimensions, estimated using MLAa, WHMC, MLAm, MAMLA, MZZS and MBPS algorithms. The grey line in each box represent the true mean of the associated component.}
	\label{fig:d10_box}
\end{figure}

We plot in Figure~\ref{fig:d10_box} the box plots for the expectations of the first three components using the algorithms discussed above. In each case, we have used $10^{2}$ realisations of the algorithms. In the case of MLAa, MLAm and MAMLA, one step of the algorithm costs one gradient evaluation, while for MZZS and MBPS, generating a candidate inter-jump time $\tau^*$ costs one gradient evaluation. For MLAa, MLAm, MAMLA, MZZS and MBPS, each realisation uses $10^7$ gradient evaluations. Furthermore, for MZZS and MBPS once a skeleton has been constructed, we generate $10^7$ samples. The number of gradient evaluations used for WHMC is $10^6$ as this takes the same runtime as the other algorithms. For each realisation of the samplers we implement a $10\%$ burn in phase. 
As we can see in Figure~\ref{fig:d10_box}, for the first component both MLAa and MLAm are biased, with MLAa  having greater variance. There is no visible bias for the mirror PDMPs and MAMLA, as they agree with WHMC. In the case of the second and third components all the samplers behave in a similar unbiased manner.

\subsection{Bayesian Inference on Dirichlet posterior}\label{numeric3} 

A distribution frequently occurring in machine learning applications \citep{mcauliffe2006nonparametric,fei2005bayesian}, particularly in the area of text modeling \cite{blei2003latent} is the Dirichlet distribution. 
A typical problem in this area is the latent Dirichlet allocation (LDA). The aim of LDA is to find topics a document belongs to, based on the words in it. In this section, we focus on sampling from a Dirichlet posterior which is a fundamental component of the full LDA model with the additional advantage that the true solution is known.  

Let $n_i$ denote the number of appearances of category $i$ where   $i=1,\ldots,d$ and let $x_i$ denote the probability of an occurrence of the category $i$. We impose a conjugate Dirichlet prior with parameter ${\alpha} = (\alpha_i)_i$ on ${x}$ with $\alpha_i >0$. Since the likelihood follows a multinomial distribution, the unnormalized posterior $\pi({x}|{n})$ is the Dirichlet distribution with parameter $({n}+ {\alpha})$ defined by 
\begin{equation*}
    \pi({x}|{n}) \propto \prod_{i=1}^{d} x_i^{n_i + \alpha_i -1}, \quad  {x}\in \Delta_d = \left\{x=(x_1,\ldots,x_d) \in \mathbb{R}^d: \ \sum_{i=1}^{d} x_{i} = 1, \ x_{i} \geq 0 \right\}.
\end{equation*}
The challenge in sampling from the Dirichlet distribution arises from the fact that the negative log density $U$ is not convex and the distribution is constrained to the simplex $\Delta_d$. The most interesting regime of the Dirichlet posterior occurs when it is sparse. Specifically, this means that most of the $n_i$'s are zero and only few of them are relatively large of the order of $\mathcal{O}(d)$ (\cite{hsieh2018mirrored}). Note that the simplex $\Delta_{d}$ is a $(d-1)$-dimensional surface; hence in order to ensure integrability of the push-forward distribution, it is more convenient to work with $\mathbb{R}^{d-1}$ dimensional vectors.  We use the barrier function defined in  Table \ref{tab:barriers} for the simplex.

For simulation, we create a synthetic dataset by generating $10^4$ data points from a $5$-dimensional multinomial distribution with a random probability vector $p$ (which is given in Section \ref{app:LDA_details}).
The parameters of the Dirichlet prior are  $\alpha_i=0.1$ for all $i$.  We evaluate the performance of MZZS, MBPS and the MZZ-SS algorithms compared to MLAa \eqref{eq:MLD}, MLAm \eqref{eq:MLAm} and MAMLA as well as their stochastic versions SMLAa and SMLAm. For MZZ-SS, as discussed in Section \ref{sec:subsampling}, we use a control variate with reference point being the mode of the push-forward $e^{-V}$, while in addition we use mini-batches of size $200$. We will measure the cost of each algorithm in terms of epochs where an epoch denotes the equivalent cost of a full pass through the data. For MZZS, MBPS, MLAa, MLAm and MAMLA, one iteration involves one gradient evaluation, which corresponds to one epoch. On the other hand, MZZ-SS, SMLAa and SMLAm use a subset of the data at every iteration and therefore we can perform several iteration for the cost of one epoch. All the algorithms are run for $2.5\times 10^5$ epochs (More details on implementation in Appendix \ref{app:LDA_details}). For MZZS, MBPS and MZZ-SS, once the skeleton is constructed, we generate $2.5 \times 10^5$ samples keeping track of the cost. 

\begin{figure}[ht]
\centering
\begin{subfigure}{0.49\linewidth}
	\includegraphics[width=\linewidth, height=5cm]{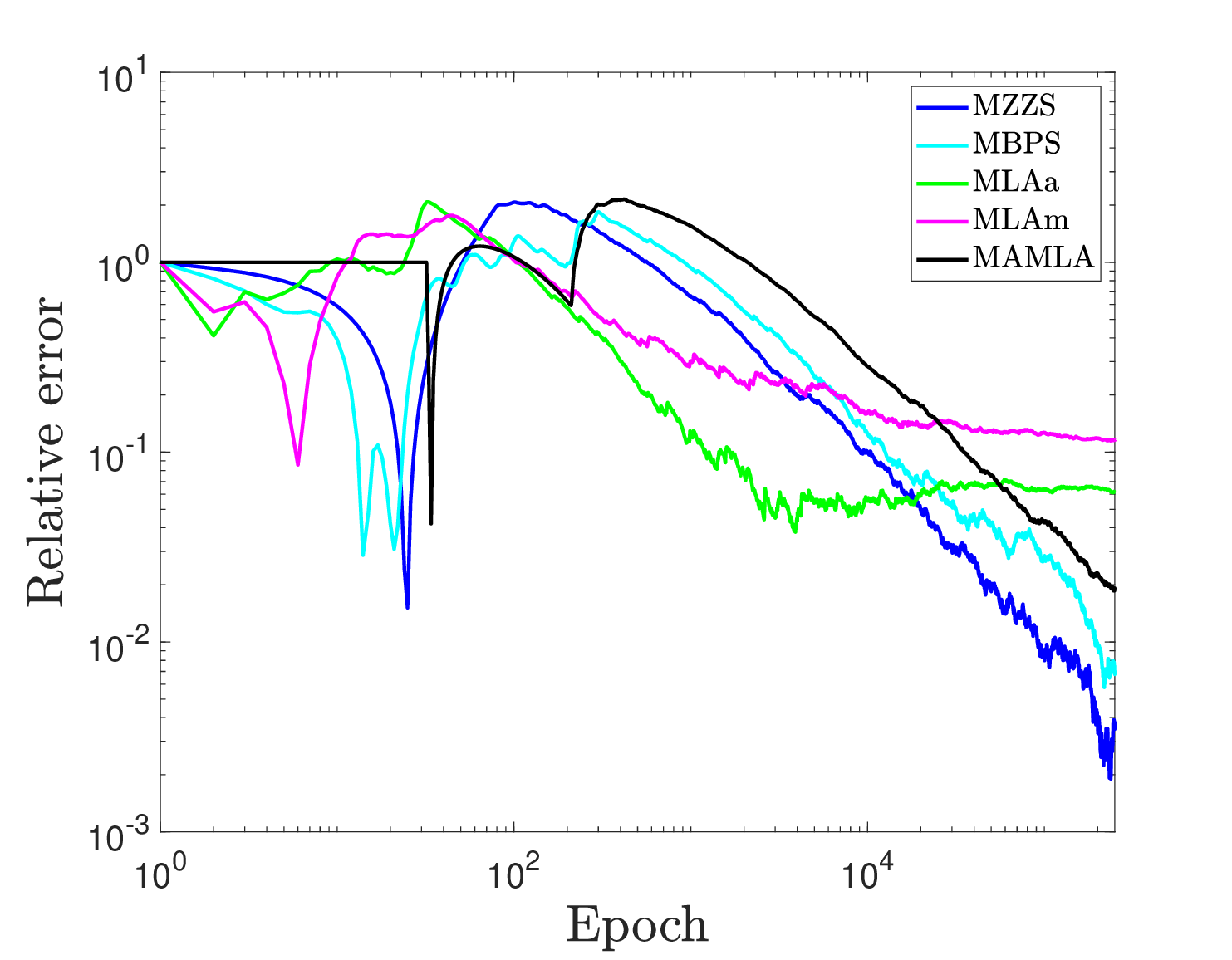}
\end{subfigure}
\begin{subfigure}{0.49\linewidth}
	\includegraphics[width=\linewidth, height=5cm]{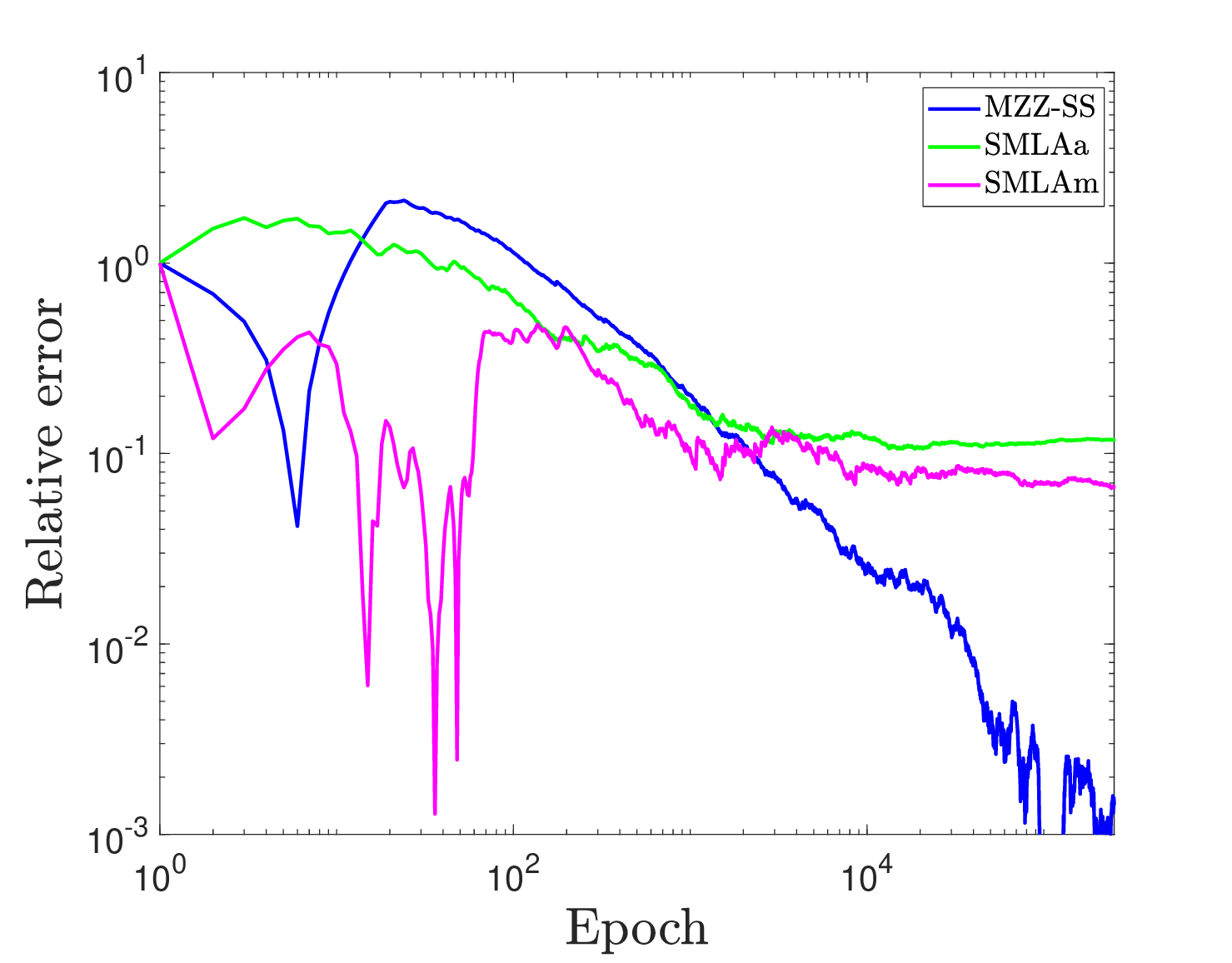}
\end{subfigure}
\caption{Plot of the running relative error in the standard deviation of the first component for samples generated by MZZS, MBPS, MLAa, MLAm,MAMLA (\textit{left}) and by MZZS-SS, SMLAa, SMLAm (\textit{right})} 
\label{fig:LDA-full}
\end{figure}

Once the samples from each algorithms are obtained, we compute the running relative error in the standard deviation of the first component as a function of the number of epochs and the results are summarised in Figure \ref{fig:LDA-full}. For the distribution $\pi$, the true standard deviation of the first component is  $\sigma_{\text{true}}=4.26\times10^{-3}.$   
As we can observe in Figure \ref{fig:LDA-full}, both MLAa and MLAm converge faster than MZZS, MBPS and MAMLA to their respective limiting behaviour but demonstrate significant bias. As expected, MZZS and MBPS have vanishing bias and converge faster than MAMLA.  SMLAa converges faster than MLAa but show higher levels of bias. In contrast, MZZ-SS converges faster than MZZS but remains unbiased. Since the constant $L_V^L$ does not exist (see details in Section \ref{app:Rel-Lip}) we plot the error for both MLAm and SMLAm with step size $\Delta t = (10L_V)^{-1}$.

\subsection{Bayesian reconstruction of quantized Gaussian processes}\label{sec:qgp} We consider a scenario in which we observe a function $f:\mathbb{R}\to\mathbb{R}$ through a quantization process. Let $f_i = f(x_i), \ i=1,\dots,n$ denote the true signal evaluated at given locations $x_i$. Instead of observing $f_i$ directly, we only observe a quantized version $y_i$ taking values in a finite set $\{ q_1, \dots, q_{N_q} \}$. Our goal is to reconstruct the values $f(x_{i})$ from these quantized observations. The quantization rule is given by  
\begin{equation}\label{eq:quantized-rule}
    y_i = q_k \quad \text{if} \quad z_k \leq f_i < z_{k+1} 
\end{equation}
where $\{z_k\}_{k=1}^{N_z}$ are the quantization thresholds. Denoting by $y = (y_1,\dots, y_n)$ and $f = (f_1,\dots, f_n)$, we aim to sample from the posterior distribution $\pi(f|y)$. 
We impose a Gaussian prior $\pi_0(f)$ on $f$ with covariance $\Sigma$, i.e. $f \sim \mathcal{N}(0,\Sigma)$ where the covariance matrix $\Sigma$ is defined through the squared exponential kernel $K$
by \begin{equation}\label{eq:kernel}
    \Sigma_{ij} = K(|x_i-x_j|) = \sigma^2 \exp \left( -\frac{|x_i-x_j|^2}{2 \eta^2} \right),
\end{equation}
where $\sigma^2$ is the variance and $\eta$ the length-scale which controls the smoothness. From the quantization rule \eqref{eq:quantized-rule}, the likelihood $p(y|f)$ of the observed data is given by 
\begin{equation*}
    p(y|f) = \prod_{i=1}^n p(y_i|f_i) 
\end{equation*}
with 
\begin{equation*}
    p(y_i|f_i) = \begin{cases}
        1 & \text{if} \quad f_i \in [z_k,z_{k+1}), \\
        0 & \text{else.} 
    \end{cases}
\end{equation*}
By Bayes' theorem, the posterior distribution is given by 
\begin{equation*}
    \pi(f|y) \propto \exp \left( -\frac{1}{2} f^\top \Sigma^{-1} f \right) \bm{1}_{\mathcal{M}}(f), \quad f\in \mathbb{R}^n 
\end{equation*}
where $\mathcal{M}$ is an $n$-dimensional hyper-rectangle determined by the data. Since the covariance matrix $\Sigma$ is highly ill-conditioned due to the smooth kernel $K$, to ensure numerical stability we regularize it  as $\Sigma_{\epsilon} = \Sigma + \epsilon I$ for a small $\epsilon >0.$ The barrier function we use here is a linear combination of the corresponding barrier functions in Table \ref{tab:barriers} for a rectangle when $a_{i}$ and $b_{i}$ are finite and a positive orthant otherwise.

For simulation similarly to \cite{pakman2014exact}, we use $n=200, \ \sigma^2 = 0.6, \ \eta^2 = 0.2$ and $\epsilon=10^{-3}$. The locations $(x_i)_{i=1}^n$ are uniformly spaced over the interval $[0,20]$. The quantization thresholds defining the set $\mathcal{M}$ are elements of $\{-\infty, -0.5,0,0.5, + \infty\}$ while the corresponding quantized values $q_i \in \{ -0.75, -0.25, 0.25, 0.75 \}$. We run WHMC  for $10^6$ iterations, while MAMLA and MZZS are run for $2 \times 10^6$ iterations in order to match the same computational time as WHMC. For MZZS, we generate $2 \times 10^6$ samples from the skeleton and a 10\% burn-in is discarded for all methods.

\begin{figure}
    \centering
   \begin{subfigure}{0.49\textwidth}
        \centering
        \includegraphics[width=\linewidth]{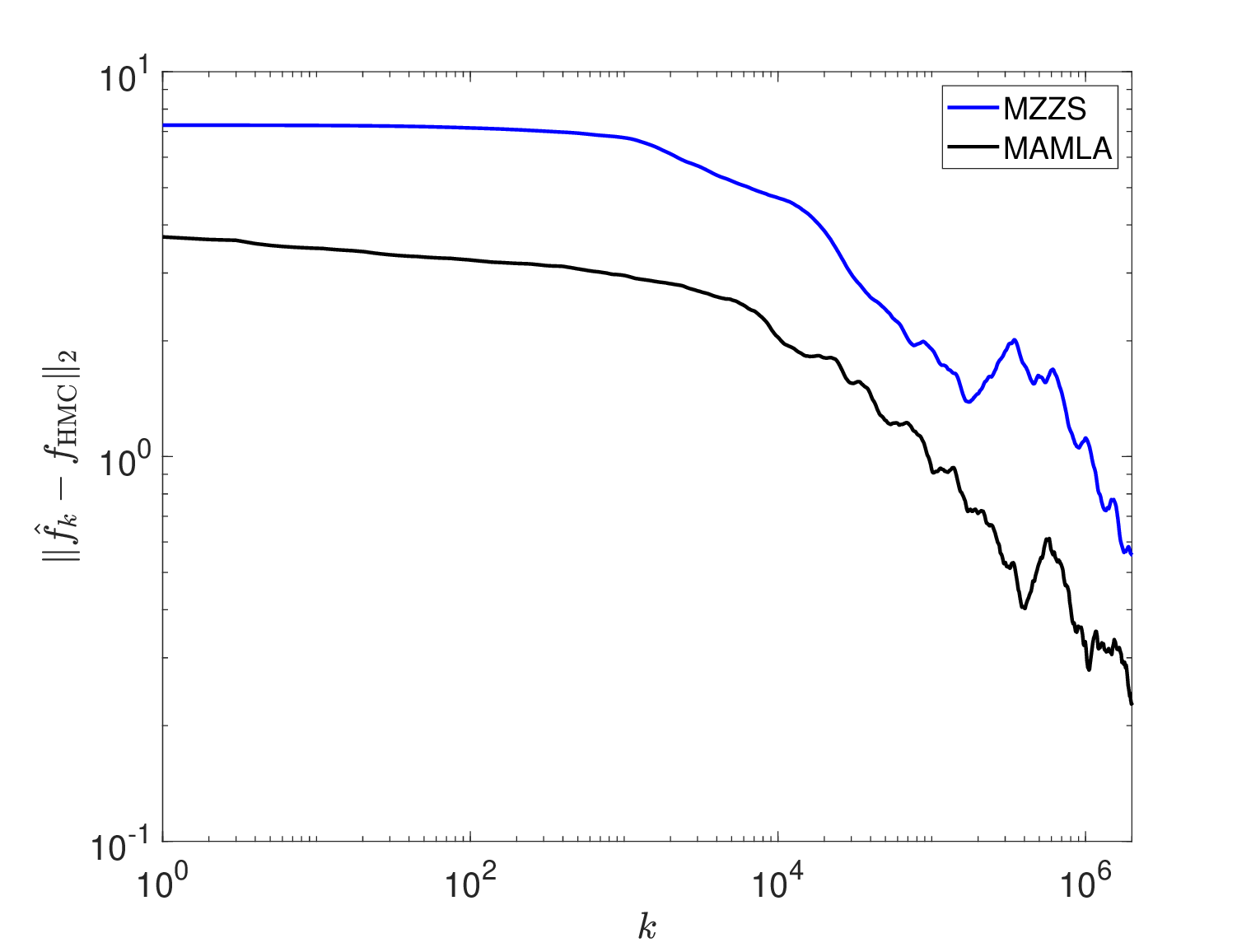}
    \end{subfigure}
    \hfill
    \begin{subfigure}{0.49\textwidth}
        \centering
        \includegraphics[width=\linewidth]{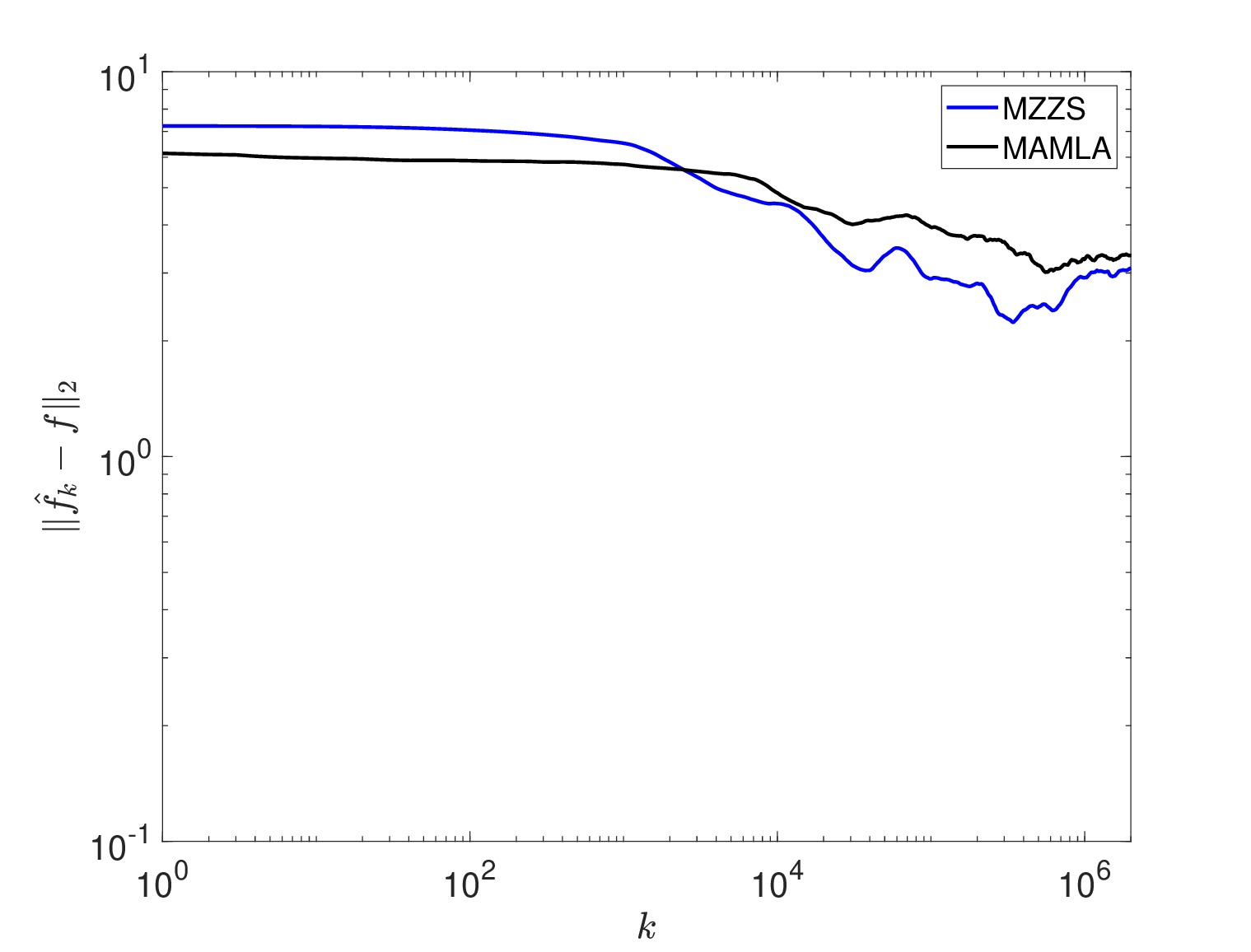}
    \end{subfigure}   
    \caption{Evolution of the $L^2$-error of the running mean of the posterior samples: comparison with  the posterior mean calculated using WHMC \textit{(left)} and with the true function $f$ \textit{(right)}.}
    \label{fig:L2_error_mzzvsmamla}
\end{figure}

For visualisation, we approximate the true signal using the posterior mean of the samples obtained for each methods and additionally, we quantified the uncertainty by plotting the 25\% and 75\% posterior quantiles for each of the methods. The results are summarised in Figure \ref{fig:qgp_reconstruction} below.  To further assess the quality of the reconstruction, we show in Figure \ref{fig:L2_error_mzzvsmamla}, the evolution of the $L^2$-error of the running mean of the posterior samples, using both the WHMC posterior mean ($f_{\text{HMC}}$) and the true signal as references. The results indicate that the reconstruction obtained with MZZS is behaving to a similar fashion  to MAMLA.   
\begin{figure}
    \centering
    \begin{subfigure}{0.49\textwidth}
        \centering
        \includegraphics[width=\linewidth]{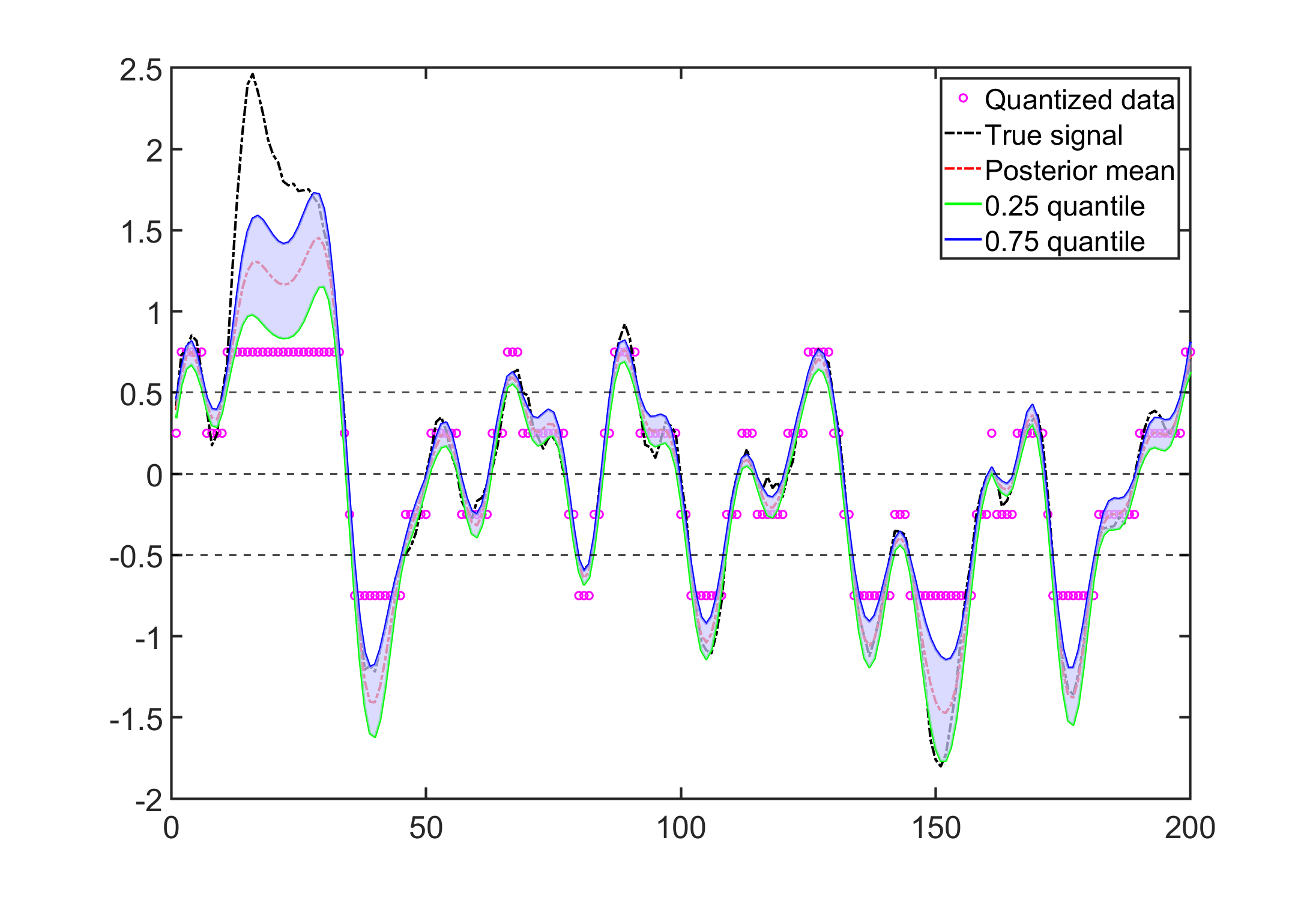}
        \caption{WHMC}
    \end{subfigure}
    \hfill
    \begin{subfigure}{0.49\textwidth}
        \centering
        \includegraphics[width=\linewidth]{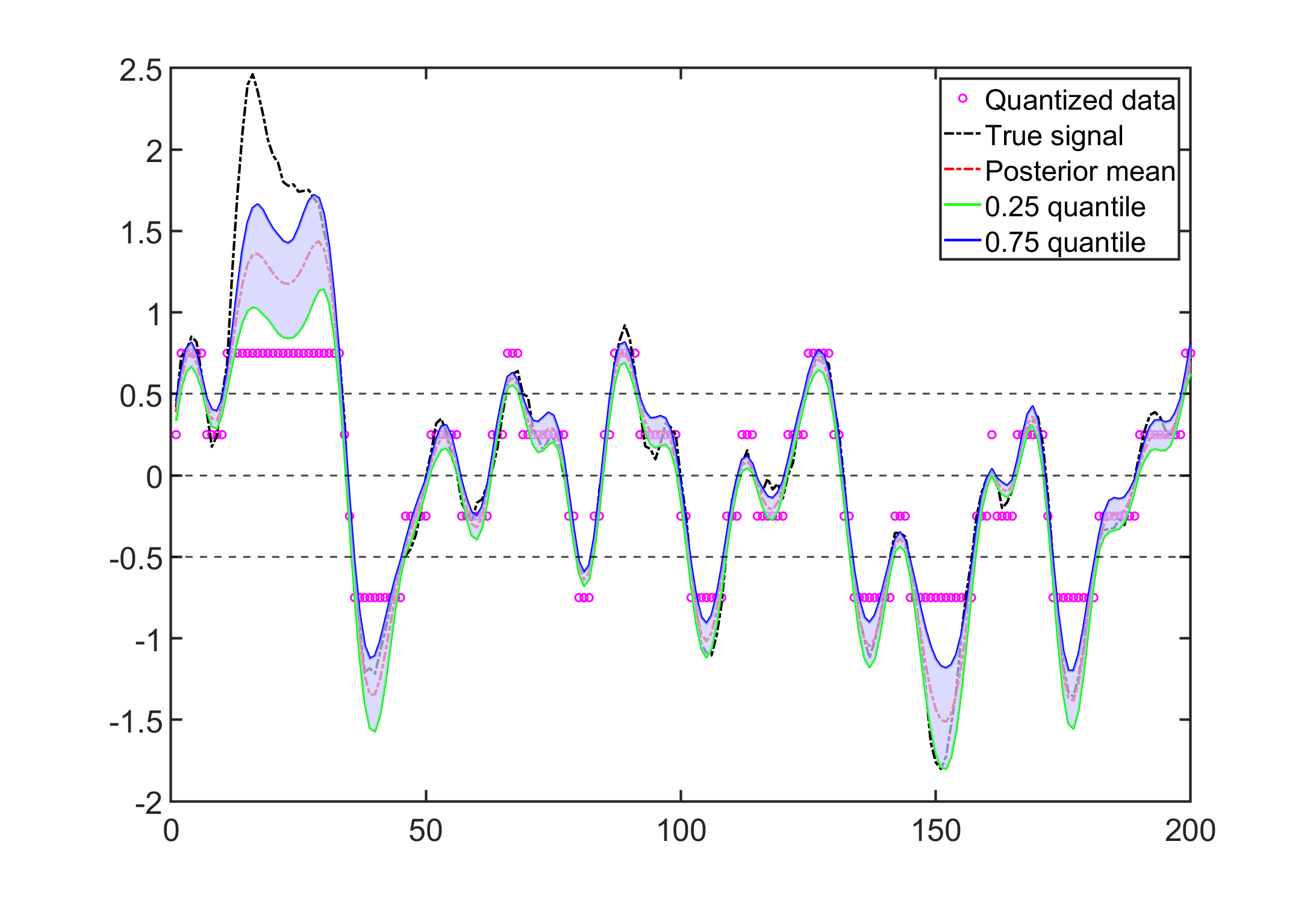}
        \caption{MAMLA}
    \end{subfigure}
    \vspace{0.5em}
    \begin{subfigure}{0.49\textwidth}
        \centering
        \includegraphics[width=\linewidth]{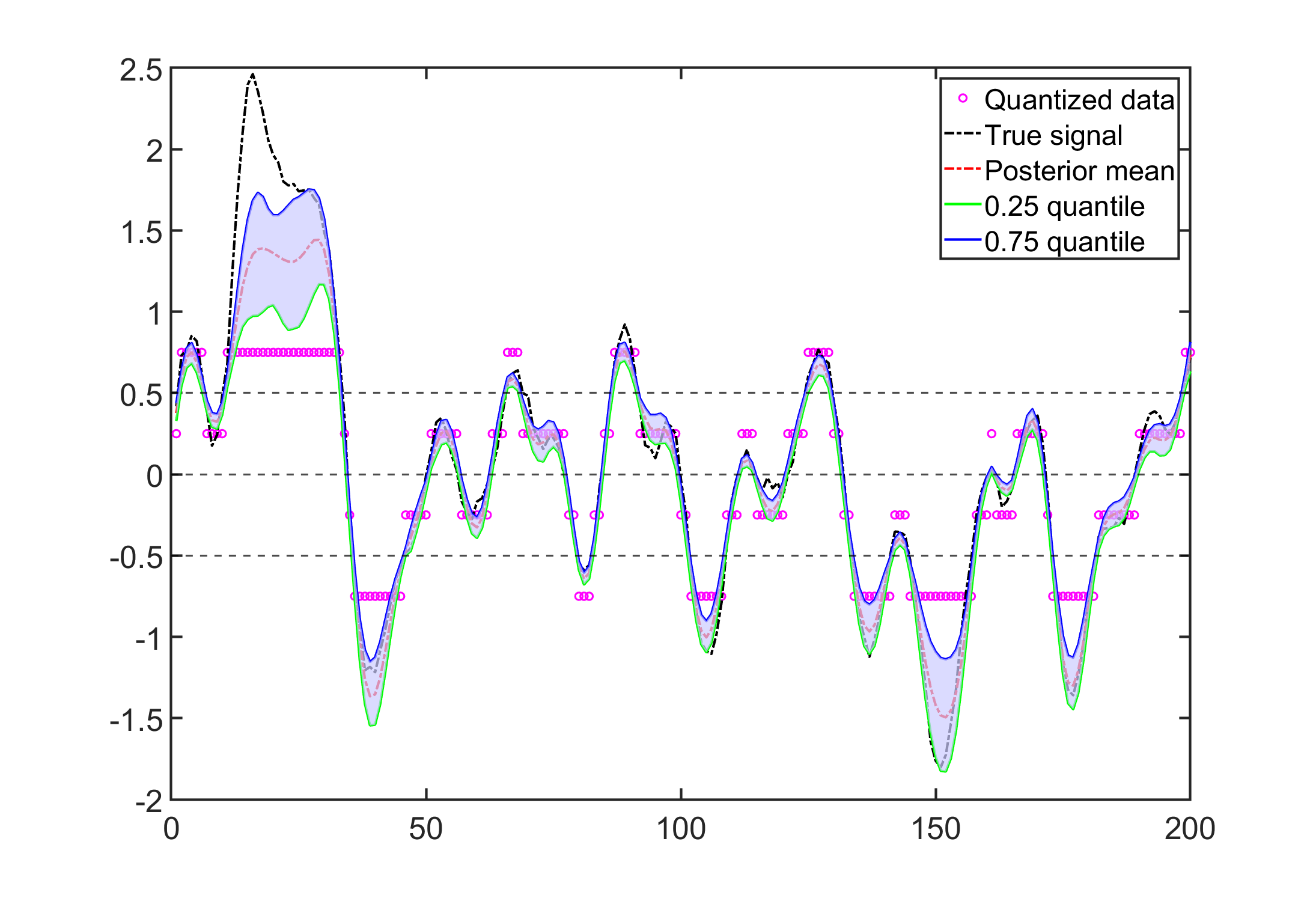}
        \caption{MZZS}
    \end{subfigure}
    \hfill
    \begin{subfigure}{0.49\textwidth}
        \centering
        \includegraphics[width=\linewidth]{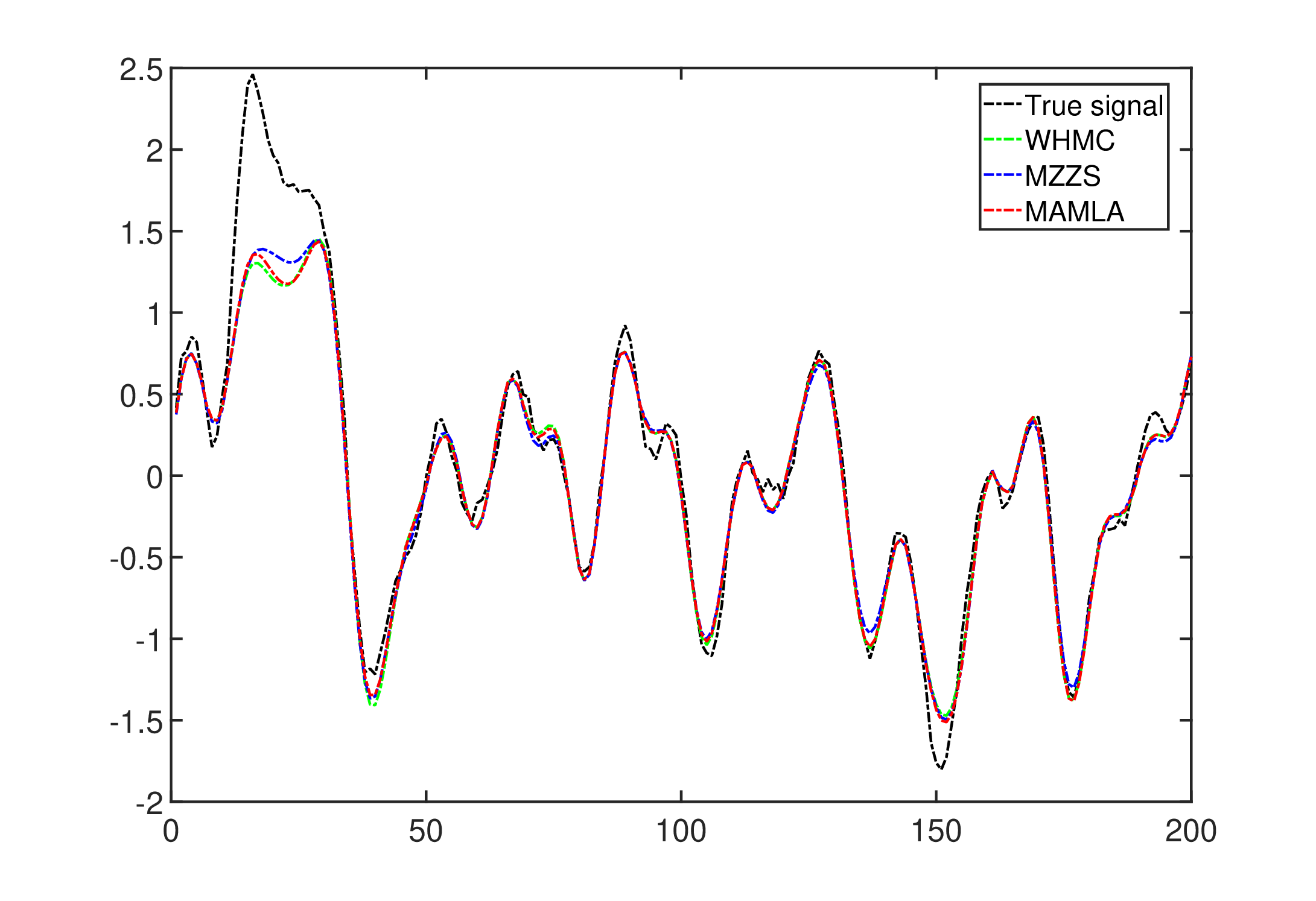}
        \caption{Posterior Mean}
    \end{subfigure}
    \caption{ayesian reconstruction of a quantized Gaussian process. The bottom-right panel shows the posterior means for each method and the true signal. 
    The remaining panels display the reconstructed signal using WHMC, MAMLA, and MZZS. 
    The shaded region corresponds 
    to the inter-quartile range (25\%–75\%).}
    \label{fig:qgp_reconstruction}
\end{figure} 

\subsection{Bayesian logistic regression on constrained domains}\label{MZZ-LogReg}  In this experiment, we consider a Bayesian logistic regression model with features $x_{j} \in \mathbb{R}^{d}$ and response variable $y_{j} \in \{0,1\}$. We impose a uniform prior on $[-a,a]^{d}$ which leads to the posterior distribution $\pi$ given by \eqref{eq:prob} with $\mathcal{M}=[-a,a]^{d}$ and 
\begin{equation*}
    U(\theta) = \sum_{j=1}^n \log(1+ \exp(\theta^\top x_j)) - y_j (\theta^\top x_j).
\end{equation*}

\begin{figure}[t]
\centering
\begin{subfigure}{0.49\linewidth}
	\includegraphics[width=\linewidth, height=5cm]{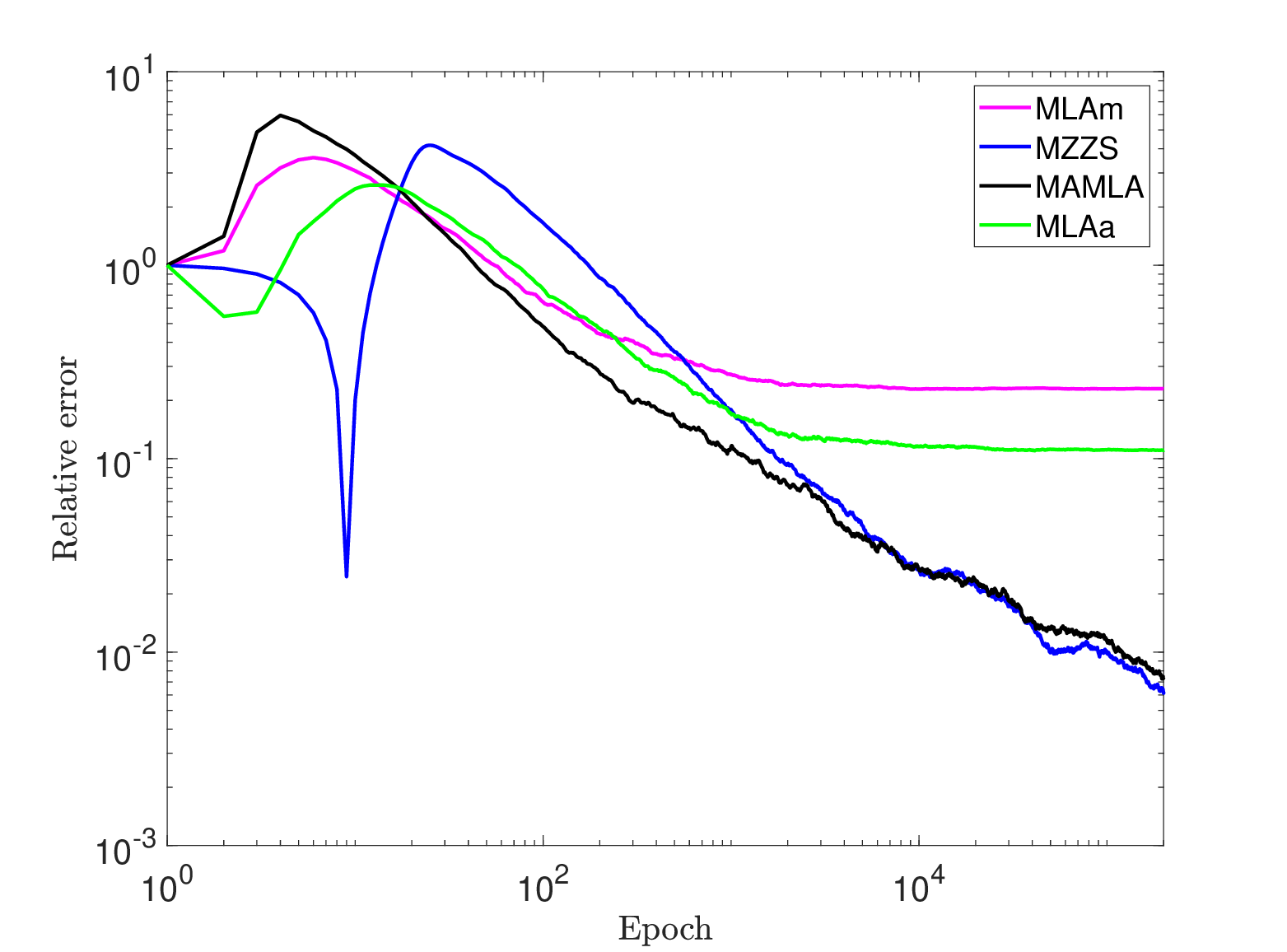}
\end{subfigure}
\begin{subfigure}{0.49\linewidth}
	\includegraphics[width=\linewidth, height=5cm]{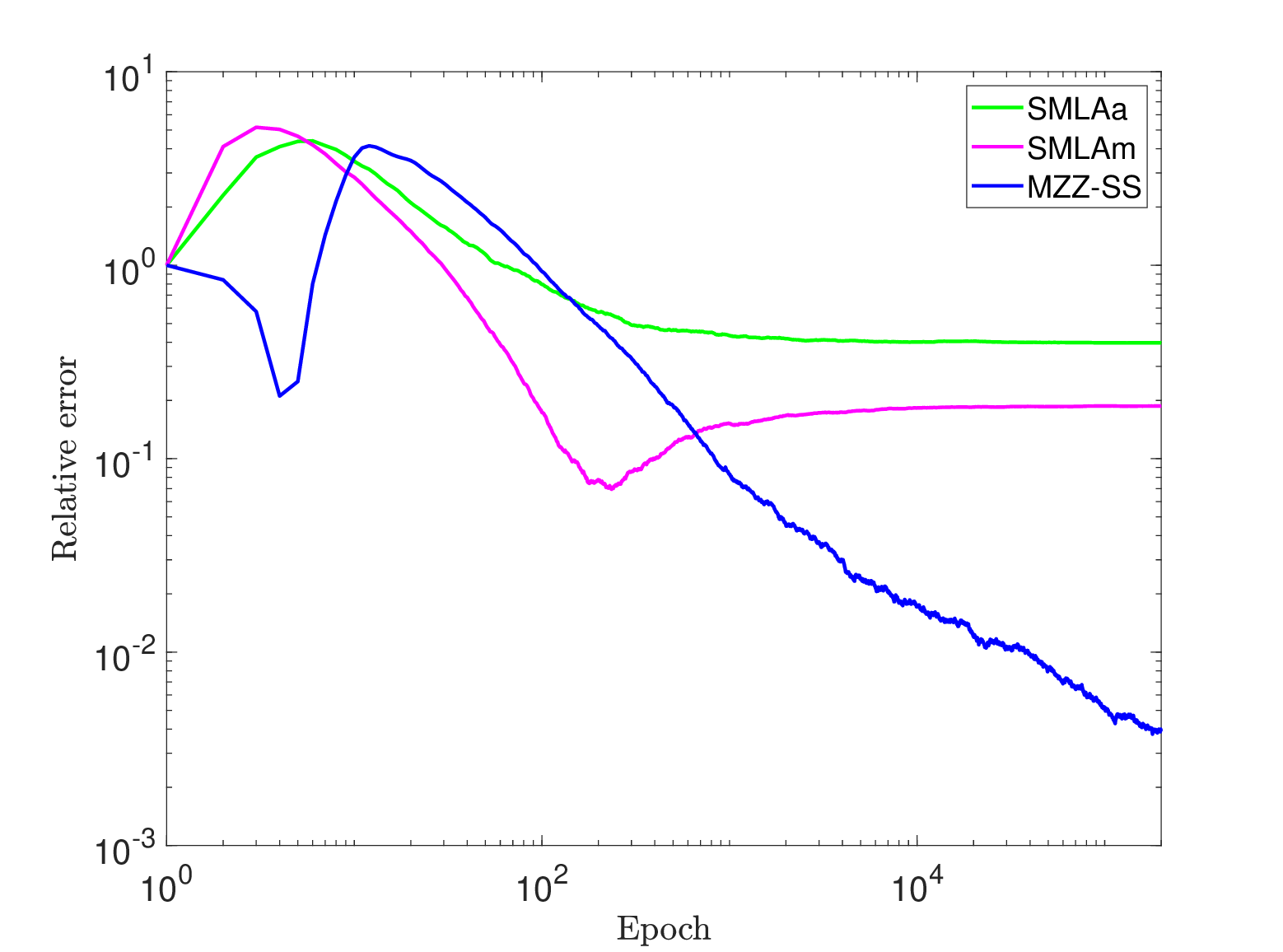}
\end{subfigure}  \\
\begin{subfigure}{0.49\linewidth}
	\includegraphics[width=\linewidth, height=5cm]{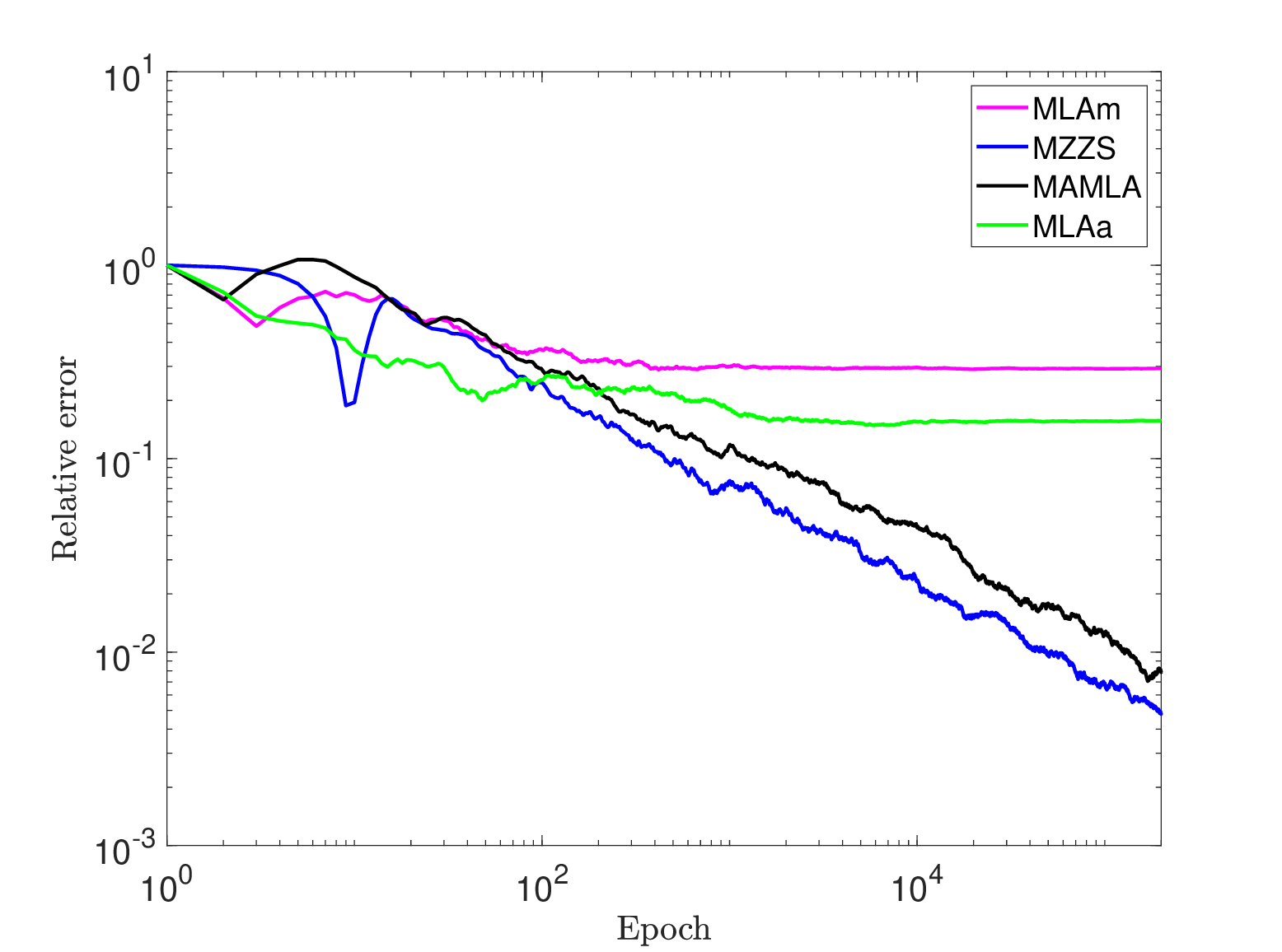}
\end{subfigure} 
\begin{subfigure}{0.49\linewidth}
	\includegraphics[width=\linewidth, height=5cm]{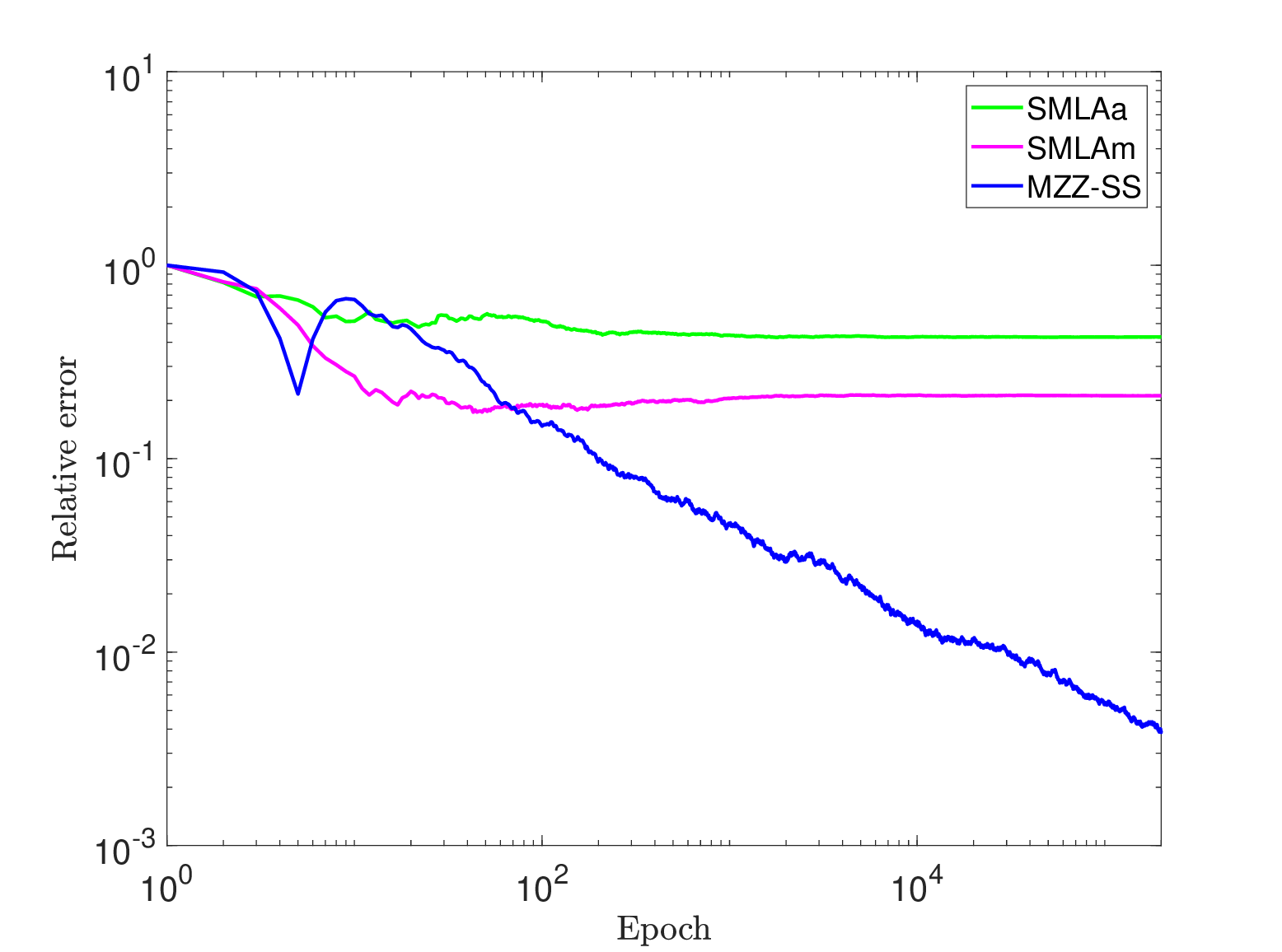}
\end{subfigure}
\caption{A plot of the running relative error of the variance for the two components of the samples generated by MLAm, MZZS, MAMLA, MLAa \textit{(left)} and SMLAm, SMLAm, MZZ-SS \textit{(right)}. \textit{Top panel: component $x_1$} and \textit{bottom panel: component $x_2$}.} 
\label{fig:const_logreg}
\end{figure}

For this experiment we use data given by the German credit dataset \cite{gc} with the features $x_{j}$ being the credit duration and age of the individual and the response $y_{j}$ being the credit outcome of an individual. We have chosen to restrict to $d=2$ as this allows for the calculation of the true posterior expectations by the means of numerical integration. The barrier function for this example is given in Table \ref{tab:barriers} for the rectangle with $a_{i}=-a, b_{i}=a$ with $a=2$. For this choice of parameter we find that the true posterior variances for each component are  $\sigma_{1,\text{true}}^{2}= 4.68 \times 10^{-3}$ and $\sigma_{2,\text{true}}^{2}= 4.31 \times 10^{-3}$.

Similarly to Section \ref{numeric3} once the samples from each algorithm have been obtained we compute the running relative error of the variance for each component. In addition, we repeat this procedure 50 times and we plot in Figure \ref{fig:const_logreg} the averaged relative error for the different algorithms. 
For sub-sampling algorithms to reduce the noise from using a stochastic gradient, we use a mini-batch of size $31$.  We observe that both MZZS and MZZ-SS outperform their SDE-based counterparts, as their relative variance error decreases to zero.


\section{Conclusion and future work}

In this paper we proposed a novel class of samplers, mirror PDMPs, which combine PDMPs with the concept of the mirror map. We demonstrated that such samplers benefit from the advantages of PDMPs, namely unbiasedness and the ability to perform exact subsampling while also respecting constraints on the probability distribution. These methods were compared to state of the art SDE-based methods such as PLMC, MYULA, MLAa and MLAm. In our experiments we saw that MZZS and MBPS outperformed the SDE-based algorithms.

In our experiments we saw a significant difference in the performance of MLAa and MLAm. The main difference between the two is that MLAm uses a diffusion matrix that is linked to a Riemannian structure on the space $\mathcal{M}$. One future direction is to adapt Mirror PDMPs to further exploit the underlying Riemannian manifold structure. 

In this paper we have concentrated on the most popular PDMPs, ZZS and BPS, however due to the nature of the mirror map it is often the case that the push-forward distribution on the dual space is heavy tailed and therefore it may be beneficial to consider other choices for the deterministic dynamics such as those introduced in \cite{vasdekis2023speed}.

Another direction for future work is to include the use of data driven mirror maps such as in \cite{Ye23}, this could be combined with an adaptive approach to preconditioning \cite{Bertazzi2020AdaptiveSF}. A final future direction concerns a general issue when working with PDMPs which is to find sharp bounds for the switching rates, this has been addressed in \cite{Corbella22,Bertazzi2022,bertazzi2023splitting,Pagani24} and these techniques can be explored in the context of mirror PDMPs. \\






\textbf{Acknowledgements.} JTD was supported by the EPSRC Centre for Doctoral Training in Mathematical Modelling, Analysis and Computation (MAC-MIGS), funded by the UK Engineering and Physical Sciences Research Council (grant EP/S023291/1), Heriot-Watt University and the University of Edinburgh. 




\newpage
\begin{appendices}


\section{Calculating Wasserstein errors}\label{app:wasserstein}

In this section, we will be explaining how we can compute the Wasserstein error for Section \ref{sec:num_ill}. The $W^1$-error between 1-dimensional probability distributions $\pi$ and $\widehat{\pi}$ can be computed using the formula 
\begin{equation}\label{eq:W1-form}
	W^1(\pi,\widehat{\pi}) = \int_0^1 |F^{-1}(q)-\widehat{F}^{-1}(q)|{d}q 
\end{equation}
where $F$ and $\widehat{F}$ are respectively the cumulative distribution function of $\pi$ and $\widehat{\pi}$. This can be estimated using samples from $\pi$ and $\widehat{\pi}$ as follows:
\begin{equation}\label{eq:app_W1error}	W^1(\pi,\widehat{\pi}) \approx  \frac{1}{N} \sum_{i=1}^N |X_{(i)}-Y_{(i)}|
\end{equation} 
where $(X_{(i)})^N_{i=1}$ and $(Y_{(i)})^N_{i=1}$ are respectively the sorted samples from $\pi$ and $\widehat{\pi}$. Motivated by this and the independence structure of the target in Section~\ref{sec:num_ill}, we use \eqref{eq:app_W1error} to calculate the $W^1$-error for each marginal and then sum these errors to obtain the total $W^1$ error.


\section{Calculating Lipschitz constants}\label{app:lips_const}
In this section, we derive the constants $L_V, \ L_V^S, \ L_V^L$ and $M_{\psi}$ which are used to determine the step sizes for the different problems considered in Sections \ref{sec:num_ill}, \ref{sec:num_ill_mzzs} and \ref{sec4}.

\subsection{Calculating constants for Section \ref{sec:num_ill}} \label{appendix_num_ill}
\textbf{Computations of $M_{\psi_1}$ and $M_{\psi_2}$.} Because the functions $x \mapsto -\log(1+x)$ and $x \mapsto -\log(1-x)$ are 1-self concordant, we deduce from \cite[Theorem 5.1.1]{nesterov2018lectures} that $M_{\psi_1} = \sqrt{2}$ and $M_{\psi_2}=2$. \\\\
\textbf{Computation of $L^L_V$ for $\psi_1$ and $\psi_2$.} For the barrier function $\psi_1$, we have for all $x \in [-1,1]^2$ 
\begin{align*}
    \|\nabla U(x)\|^2_{\nabla^2\psi_1(x)^{-1}} & = \frac{(1-x_1^2)^2 x_1^2}{1+x_1^2} + \frac{1}{\epsilon^2} \frac{(1-x_2^2)x_2^2}{1+x_2^2} \\
    & \leq \max_{y \in [-1,1]} \frac{(1-y^2)^2 y^2}{1+y^2} \left( 1 + \frac{1}{\epsilon^2}\right) \\
    & \leq 0.12 \left( 1 + \frac{1}{\epsilon^2}\right) =: (L^L_V)^2. 
\end{align*}
Similarly for the barrier function $\psi_2$, we have for all $x \in [-1,1]^2$
\begin{align*}
    \|\nabla U(x)\|^2_{\nabla^2\psi_2(x)^{-1}} & = \frac{2(1-x_1^2)^2 x_1^2}{1+x_1^2+(1-x_1^2)^2} + \frac{1}{\epsilon^2} \frac{2(1-x_2^2)x_2^2}{1+x_2^2+ \frac{1}{\epsilon}(1-x_2^2)^2} \\
    & \leq \frac{2}{10} \left( 1 + \frac{1}{\epsilon^2}\right) =: (L^L_V)^2
\end{align*} \\
\textbf{Computation of $L^S_V$ for $\psi_1$ and $\psi_2$.} We are looking for $L^S_V$ such that $\nabla^2 U(x) \prec L^S_V \nabla^2 \psi_i(x), \ i=1,2.$ For all $v \in \mathbb{R}^2$, we have 
\begin{align*}
   & \langle \nabla^2 U(x)v,v \rangle  = v_1^2 + \frac{1}{\epsilon}v_2^2 \leq \max \left( 1, \frac{1}{\epsilon}  \right) \|v\|^2_2 \\
   & \langle \nabla^2 \psi_1(x)v,v \rangle  = \frac{(1+x_1^2)}{(1-x_1^2)^2}v_1^2 + \frac{(1+x_2^2)}{(1-x_2^2)^2}v_2^2 \geq \|v\|_2^2.
\end{align*}
Then for the barrier function $\psi_1$ $$L^S_V = \max \left( 1, \frac{1}{\epsilon} \right).$$ 
Using the fact 
\begin{align*}
     \langle \nabla^2 \psi_2(x)v,v \rangle & = \frac{1}{2} \langle \nabla^2 \psi_1(x)v,v \rangle + \frac{1}{2} \langle \Sigma^{-1} v,v \rangle  \geq \frac{1}{2} \|v\|_2^2,  
\end{align*}
we deduce that for the barrier function $\psi_2$, we have 
$$L^S_V = \max \left( 2, \frac{2}{\epsilon} \right).$$ 
\subsection{Calculating $L_V$ for MZZS in Section \ref{sec:num_ill_mzzs}.} Given that 
\begin{equation*}
    \nabla \psi_3(x) = \left( \frac{x_1}{2(1-x_1^2)}+ \frac{x_1}{2}, \frac{x_2}{2(1-x_2^2)}+ \epsilon^{-1/2}\frac{x_2}{2} \right),
\end{equation*}
we find $\nabla \psi_3^*(\zeta)$ by solving the equation $\nabla \psi_3(x)=\zeta$ for $x$. This is equivalent to solve the cubic equation  
\begin{equation*}
    -c_ix_i^3 + 2 \zeta_i x_i^2 + (1+c_i)x_i -2 \zeta_i = 0 
\end{equation*}
for $x_i$ as a function of $\zeta_i$ where $(c_1,c_2) = (1,\epsilon^{-1/2}).$ This will defines $x = \nabla \psi^*_3(\zeta)$ implicitly. Using the expression of $U$ and the fact that $\nabla \psi_3 \circ \nabla^*_3(\zeta) = \zeta,$ we found the expression of $V$ given by 
\begin{equation*}
    V(\zeta) = \frac{1}{2} \left( x_1^2 + \frac{x_2^2}{\epsilon} \right) + \sum_{i=1}^2 \log \left[ \frac{1+x_i^2}{2(1-x_i^2)^2} + \frac{1}{2}c_i \right], \quad x = \nabla \psi_3^*(\zeta). 
\end{equation*}
The Lipschitz constant $L_V$ is therefore given by 
\begin{equation*}
    L_V = \sup_{\zeta \in \mathbb{R}^2} \|\nabla^2 V(\zeta)\|_2 = \max_{i=1,2} \sup_{x_i \in (-1,1)} \left| \frac{\partial^2 V}{\partial \zeta_i^2} \right|
\end{equation*}
which we derive to be 
\begin{equation*}
    L_V = \max \left( 1, \frac{4}{(1+\epsilon^{1/2})^2} \right).
\end{equation*}

\subsection{Calculating constants for Section \ref{numeric2}}\label{appendix:gamma}
\textbf{Computation of $L_V$.} For our barrier function, the Fenchel conjugate $\nabla\psi^*$ is defined by $$\nabla \psi^*(\zeta) = \frac{\zeta+ \sqrt{\zeta^2+4}}{2}.$$ The Hessian of $\psi^*$ is the diagonal matrix $\nabla^2 \psi^*$ given by 
\begin{equation*}
(\nabla^2 \psi^*(\zeta))_{ij}= \begin{cases}
\frac{1}{2}+ \frac{\zeta_i}{2 \sqrt{\zeta_i^2+4}} & \text{if}  \ i=j \\
0 & \text{else} 
\end{cases}
\end{equation*}
Therefore, the negative log-density of the push-forward distribution is given by 
\begin{align*}
V(\zeta) & = \sum_{i=1}^d \beta_i \left( \frac{\zeta_i+ \sqrt{\zeta_i^2+4}}{2} \right)- \alpha_i \log \left( \frac{\zeta_i+ \sqrt{\zeta_i^2+4}}{2} \right) + \log\left(\sqrt{\zeta_i^2+4}\right),
\end{align*}
and the following bound holds 
\begin{align*}
|\partial^2_{ij} V(\zeta)| & \leq \delta_{ij} \left\{\left| \frac{4-\zeta_i^2}{(\zeta_i^2+4)^2}\right| + \alpha_i \left| \frac{\zeta_i}{(\zeta_i^2+4)^{3/2}} \right| + \beta_i \frac{2}{(\zeta_i^2+4)^{3/2}}   \right\}
\leq \delta_{ij}\left(\frac{1}{4} + \frac{\alpha_i}{10} + \frac{\beta_i}{4}\right). 
\end{align*}
Then $\nabla V$ is Lipschitz continuous with Lipschitz constant $\displaystyle L_V = \frac{1}{4}+ \max_{1 \leq i \leq d} \left[ \frac{\alpha_i}{10} + \frac{ \beta_i}{4}  \right]  $

\textbf{Computation of $M_{\psi}$}. The barrier $\psi$  is 1-self concordant as the sum of self concordant functions \cite[Theorem 5.1.1]{nesterov2018lectures}, hence $M_{\psi}=1$. 

\textbf{Computation of $L_V^L$.} For all $x \in (0,+\infty)^d$, we have 
\begin{align*}
\langle \nabla^2\psi(x)^{-1} \nabla U(x), \nabla U(x) \rangle & = \sum_{i=1}^d \frac{(\beta_i x_i- (\alpha_i-1))^2}{1+x_i^2} \leq \sum_{i=1}^d \max \{\beta_i^2, (\alpha_i-1)^2\} =: (L_V^{L})^2. 
\end{align*}

\textbf{Computation of $L_V^S$.} We require $\nabla^2 U(x) \preceq L^S_V \nabla^2\psi(x)$ for all $x=(x_1,\dots,x_d)$ with $x_i > 0$, which is equivalent in our case to 
\begin{equation*}
   \sum_{i=1}^d \frac{\alpha_i-1}{x_i^2} \leq L_V^S \sum_{i=1}^d\left( 1 + \frac{1}{x_i^2} \right).
\end{equation*}
This holds if $\displaystyle L^S_V = \max_{1\leq i \leq d} \ |\alpha_i-1|.$

\subsection{Calculating constants for Section \ref{numeric1}}\label{appendix_trunc_gauss}

\textbf{Computation of $L_V$.}  For our barrier function $\psi$,  $V$ is given by $ V(\zeta) = V_1(\zeta) + V_2(\zeta)$ where 
\begin{equation*}
    V_1(\zeta) = U \circ \nabla \psi^*(\zeta), \quad V_2(\zeta) = -\log \det \nabla^2 \psi^*(\zeta) 
\end{equation*}
with $\nabla \psi^*$ defined by 
\begin{equation*}
    \nabla \psi^*(\zeta) = \left( \frac{a_i+b_i}{2} + \frac{(b_i-a_i)^2}{2\left(\sqrt{(b_i-a_i)^2\zeta_i^2 +4}+2\right)} \right)_{i=1}^d.
\end{equation*}
From the expression of $\nabla \psi^*$, we deduce that 
\begin{equation}\label{eq:g2phis_trunc_gauss}
  \nabla^2 \psi^*(\zeta)_{ij} = \begin{cases}
      \frac{(b_i-a_i)^2}{s(\zeta_i) (s(\zeta_i)+2) } & \text{if} \quad i=j \\
      0 & \text{else} 
  \end{cases}  
\end{equation}
with $s(t) = \sqrt{(b_i-a_i)^2 t^2 +4}$. Using \eqref{eq:g2phis_trunc_gauss}, we deduce that 
\begin{align*}
    V_2(\zeta) & = -\log \det \nabla^2 \psi^*(\zeta) = -\sum_{i=1}^d \log \left( \frac{(b_i-a_i)^2}{s(\zeta_i)(s(\zeta_i)+2)} \right) \\
    & = \sum_{i=1}^d \left\{  \log(s(\zeta_i)+2) + \log s(\zeta_i) -2 \log(b_i-a_i) \right\}
\end{align*}
and hence 
\begin{equation*}
    \nabla V_2(\zeta) = \left( (b_i-a_i)^2 \zeta_i \left[ \frac{1}{s(\zeta_i)^2} + \frac{1}{s(\zeta_i)(s(\zeta_i)+2)} \right] \right)_{i=1}^d.
\end{equation*}
A simple calculation show that the function $\varphi_i: t \mapsto \frac{d}{dt}\partial_i V_2(t)$ is given by 
\begin{equation*}
    \varphi_i(t) = -2(b_i-a_i)^4 t^2 \left[ \frac{1}{s(t)^4} + \frac{1+s(t)}{s(t)(s(t)+2)} \right] + (b_i-a_i)^2 \left[ \frac{1}{s(t)^2} + \frac{1}{s(t)(s(t)+2)} \right]
\end{equation*}
which is even and satisfy  $|\varphi_i(t)| \leq \varphi_i(0) = \frac{3}{8}(b_i-a_i)^2$. Then $\nabla V_2$ is $L_{V_2}$-Lipschitz with 
\begin{equation*}
    L_{V_2} = \frac{3}{8}\|b-a\|^2_{\infty}. 
\end{equation*}
The gradient of $V_1$ is given by $\nabla V_1(\zeta) = \nabla^2 \psi^*(\zeta) \Sigma^{-1} \nabla \psi^*(\zeta)$. For $\zeta_1, \zeta_2 \in \mathbb{R}^d$, we have $\|\nabla V_1(\zeta_1)- \nabla V_1(\zeta_2)\|_2 \leq AB + CD$ with 
\begin{align*}
    & A = \|\nabla^2 \psi^*(\zeta_1)\|_2,  & B = \|\Sigma^{-1}(\nabla\psi^*(\zeta_1) - \nabla\psi^*(\zeta_2)) \|_2 \\
    & C = \|\Sigma^{-1}\nabla\psi^*(\zeta_2)\|_2,  & D = \|\nabla^2 \psi^*(\zeta_1)- \nabla^2 \psi^*(\zeta_2)\|_2.
\end{align*}
From \eqref{eq:g2phis_trunc_gauss}, we deduce that 
\begin{align*}
    & A = \max_{1 \leq i \leq d} \frac{(b_i-a_i)^2}{s(\zeta_i)(s(\zeta_i)+2)} = \max_{1 \leq i \leq d} \frac{(b_i-a_i)^2}{8} =: \frac{3}{8}\|b-a\|^2_{\infty}, \\
   & B  \leq \|\Sigma^{-1}\|_2 \|\nabla \psi^*(\zeta_1)- \nabla \psi^*(\zeta_2)\|_2 
    \leq \|\Sigma^{-1}\|_2 \frac{3}{8}\|b-a\|^2_{\infty} \|\zeta_1-\zeta_2\|_2 \\
    & C \leq \|\Sigma^{-1}\|_2 \|\nabla\psi^*(\zeta_2)\|_2 \leq \|\Sigma^{-1}\|_2 \sqrt{\sum_{i=1}^d \left( \frac{a_i+b_i}{2} + \frac{(b_i-a_i)^2}{8} \right)^2 }.
\end{align*} 
For the term $D = \|\nabla^2 \psi^*(\zeta_1)- \nabla^2 \psi^*(\zeta_2)\|_2$, an easy computation show that the map $\varphi_i: t \mapsto \frac{d}{dt} \nabla^2 \psi^*(t)_{ii}$ is given by 
\begin{equation*}
    \varphi_i(t) = \frac{2(b_i-a_i)^4t(1+s(t))}{s(t)^3(s(t)+2)^2} 
\end{equation*}
and satisfy $|\varphi_i(t)| \leq \frac{(b_i-a_i)^3}{8}$ and we have 
\begin{equation*}
    D \leq \frac{\|b-a\|^3_{\infty}}{8} \|\zeta_1-\zeta_2\|_2.
\end{equation*}
Combining the previous computations gives
\begin{align*}
   L_V = \|\Sigma^{-1}\|_2 \frac{\|b-a\|^3_{\infty}}{8} \left[ \frac{9}{8}\|b-a\|_{\infty} + \sqrt{\sum_{i=1}^d \left( \frac{a_i+b_i}{2} + \frac{(b_i-a_i)^2}{8} \right)^2 } \right] 
    + \frac{3}{8} \|b-a\|^2_{\infty}. 
\end{align*}

\textbf{Computation of $M_{\psi}$.} The barrier  function  is 1-self concordant since the functions $x \mapsto -\log(x-a)$ and $x \mapsto -\log(b-x)$ are 1 self-concordant \cite[Theorems 5.1.1 and 5.1.4]{nesterov2018lectures}.\\

\textbf{Computation of $L_V^L$}. Since $\Sigma$ is symmetric, then by the spectral theorem, we can express any vector $x \in \mathbb{R}^d$ as $x = \sum_{i=1}^d \gamma_i \hat{e}_i$ where $\gamma_i \in \mathbb{R}$ and $(\hat{e}_i)_{i=1}^d$ is an eigenbasis for $\Sigma$ with eigenvalues $\lambda_i$. Since $\nabla U(x) = \Sigma^{-1}x$, we have 
\begin{align*}
\|\nabla U(x)\|^2_{\nabla^2\psi(x)^{-1}}
& = \sum_{i=1}^d \sum_{j=1}^d \langle \nabla^2\psi(x)^{-1} \lambda_i^{-1}\gamma_i \hat{e}_i, \lambda_j^{-1}\gamma_j \hat{e}_j\rangle  \\ 
& \leq \left( \max_{1\leq i \leq d} \lambda_i^{-1} \right)^2  \langle \nabla^2\psi(x)^{-1} x, x \rangle. 
\end{align*}
Let $(x_i)_{i=1}^d$ be the coordinate of $x$ in the canonical basis $(e_i)_{i=1}^d$ of $\mathbb{R}^d$. Using the fact that $\nabla^2\psi(x)^{-1}$ is a diagonal matrix, we have 
\begin{align*}
 \|\nabla U(x)\|^2_{\nabla^2\psi(x)^{-1}}   & \leq \left( \max_{1\leq i \leq d} \lambda_i^{-1} \right)^2 \sum_{i=1}^d (\partial^2_{ii}\psi(x))^{-1} x_i^2.
\end{align*}
Since $\psi$ is separable, the function $h_i(x_i) = (\partial^2_{ii}\psi(x))^{-1} x_i^2$ is a function of only $x_i$ and hence 
\begin{equation}
    \|\nabla U(x)\|^2_{\nabla^2\psi(x)^{-1}}  \leq \|\Sigma^{-1}\|^2_2 \sum_{i=1}^d h_i(x_i)
\end{equation}
with $\|\Sigma^{-1}\|_2$ being the spectral norm of $\Sigma^{-1}$. Using the expression of $\psi$, the function $h_i$ introduced above can be written as
\begin{equation*}
h_i(x_i) = \frac{x_i^2 (a_i-x_i)^2 (b_i-x_i)^2}{(a_i-x_i)^2+(b_i-x_i)^2}.
\end{equation*} 
The function $h_i$ reaches its supremum on $(a_i,b_i)$ at $(a_i+b_i)/2$ and $h_i\left( \frac{a_i+b_i}{2} \right) = \frac{(b_i-a_i)^2 (b_i+a_i)^2}{32}.$
Therefore, we have  $\displaystyle h_i(x_i) \leq \max_{1\leq i\leq d} \frac{(b_i-a_i)^2 (b_i+a_i)^2}{32}$ and 
\begin{equation*}
\|\nabla U(x)\|_{\nabla^2\psi(x)^{-1}} \leq \|\Sigma^{-1}\|_2 \sqrt{\sum_{i=1}^d \frac{(b_i-a_i)^2(b_i+a_i)^2}{32}}=:L_V^L.
\end{equation*}
\textbf{Computation of $L_V^S$.} For relative smoothness, we want to find $L_V^S$ such that $\langle \nabla^2 U(x)v, v \rangle  \leq L_V^S \langle \nabla^2 \psi(x)v,v \rangle$ for all $v \in \mathbb{R}^d$.
We have 
\begin{align*}
& \langle \nabla^2 U(x)v,v \rangle \leq \|\nabla^2 U(x)\|_2 \|v\|^2 = \|\Sigma^{-1}\|_2 \|v\|^2, 
\end{align*}
where $\lVert \cdot \rVert_2$ denotes the spectral norm. 
Additionally, a simple calculation allows us to show that 
\begin{equation*}
\langle \nabla^2 \psi(x)v,v \rangle \geq  \frac{8}{\|b-a\|^2_{\infty}}\|v\|^2. 
\end{equation*} 
This leads to   
\begin{equation*}
L_V^S = \frac{\|\Sigma^{-1}\|_2 \|b-a\|^2_{\infty}}{8}. 
\end{equation*}

\subsection{ Calculating constants for Section \ref{numeric3}} \label{app:LDA_details}
For the simulation in Section \ref{numeric3}, the data is generated using the probability vector $$  p=(0.2401,0.2669,0.0374,0.2692,0.1864).$$ 

\noindent \textbf{Implementation details of MZZ-SS.}
Since $U$ is given by \eqref{eq:sum_form}, the function $V$ given by \eqref{pushforward-potential} can be written in the form \eqref{eq:sum_form} with $U$ and $U^j$ replaced respectively by $V$ and $V^j$ where
$$
V^j(\zeta) = - \left( K m^j + \alpha \right)\zeta + (N_d + \Gamma_d) \psi^*(\zeta).
$$
Here $K$ is the number of batches, $m^j_i$ is the number of occurrences of category $i$ in the batch $j$ and

$$
n_i = \sum_{j=1}^K m^j_i, \quad m^j = \left(m^j_i\right)_{i=1}^d, \quad  N_d = \sum_{i=1}^{d}n_i, \quad \Gamma_d = \sum_{i=1}^{d} \alpha_{i}.
$$
Note that unlike $U^j$ the function $V^j$ is strongly convex. 
Analogously to the setting of  Section \ref{sec:subsampling}, for MZZ-SS we use the control variate approach with a bound of the  form \eqref{eq:ZZ-SS bound} given by 
\begin{align}\label{eq:SMZZS_bound}
    \widetilde{M}_i(t) := (v_i \partial_i V(\bar{\zeta}))^+ + L_{V,p}\|\zeta-\bar{\zeta}\|_p + t L_{V,p} \|v\|_p
\end{align}
with $p=\infty$, $L_V=L_{V,\infty}=(N_d+\Gamma_d)$ the Lipschitz constant of $\nabla V^j$ and $\bar{\zeta}$ being the mode of the push-forward given by $$\bar{\zeta} = \nabla\psi\left( \frac{1}{N_d+\Gamma_d} \left(n + \alpha\right) \right).$$ For all the algorithms the initial position is given by the centre of the simplex, i.e. $x_0=(1/d,\ldots,1/d)$. For MZZS, MBPS and MZZ-SS the initial velocity is a random sample from the velocity marginal of the invariant distribution, i.e. $v_0\sim \mathrm{Unif}(\{\pm1\}^d)$ for MZZS and MZZ-SS while $v_0 \sim \mathrm{Unif}(\mathbb{S}^d)$ for MBPS. \\
\textbf{Calculation of $L_V$.} Using our barrier function we deduce that 
\begin{equation*}
    \nabla\psi^*(\zeta) = \left( \frac{e^{\zeta_i}}{1+ \sum_{k=1}^{d-1} e^{\zeta_k}} \right)_{i=1}^d. 
\end{equation*}
Observe that  
\begin{align*}
    V(\zeta)  = -\sum_{i=1}^{d-1} (n_i+\alpha_i)\zeta_i + (N_d+ \Gamma_d) \psi^*(\zeta). 
\end{align*}
Finally, by setting $\bm{n} = (n_1, \dots,n_{d-1})^\top$ and $\bm{\alpha} = (\alpha_1,\dots,\alpha_{d-1})^\top$, we have 
\begin{equation*}
    \nabla V(\zeta) = -(\bm{n}+ \bm \alpha) + (N_d + \Gamma_d) \nabla \psi^*(\zeta),
\end{equation*}
which is $L_V = (N_d+\Gamma_d)$-Lipschitz. \\
\textbf{Computation of $M_\psi$.} Because the function $g: x \mapsto x\log x$ for $x \in (0,1)$ satisfy $$\frac{|g'''(x)|}{|g''(x)|^{3/2}} = \frac{1}{\sqrt{x}} \longrightarrow \infty \quad \text{as} \ x \to 0,$$
there is no $M_g < \infty$ such that $|g'''(x)| \leq 2 M_g |g''(x)|^{3/2}$ and hence $\psi$ is not self concordant, i.e. $M_\psi = + \infty. $ 
\\\\
\textbf{Calculation of $L_V^L$.}\label{app:Rel-Lip}
For the Bayesian inference experiment with Dirichlet distribution of Section \ref{numeric3}, the function $U$ is defined as 
\begin{equation}\label{eq:pot_dirichlet}
    U(x) = -\sum_{i=1}^{d-1} (n_i+\alpha_i-1)\log (x_i) - (n_d+\alpha_d-1)\log(x_d) 
\end{equation}
where $x_d = 1-\sum_{i=1}^{d-1}x_i$. We deduce that 
\begin{equation*}
    \nabla U(x) = \left( -\frac{n_i+\alpha_i-1}{x_i} + \frac{n_d+\alpha_d-1}{x_d} \right)_{i=1}^{d-1}
\end{equation*}
and using our barrier function
\begin{equation*}
    \partial^2_{ij}\psi(x) = \begin{cases}
        \frac{1}{x_i} + \frac{1}{x_d} & \mathrm{if} \ i=j \\
        \frac{1}{x_d} & \mathrm{if} \ i \neq j.
    \end{cases}
\end{equation*}
This leads to $\nabla^2\psi(x) = \mathrm{diag}(1/x_1,\cdots,1/x_{d-1})+ \frac{1}{x_d}\mathbf{1}\mathbf{1}^\top$, where $\mathbf{1}$ is the $(d-1)$-dimensional vector with all components being 1. The Sherman-Morrison formula allows us to get the inverse of $\nabla^2\psi(x)$ given by $\nabla^2\psi(x)^{-1} = \mathrm{diag}(x_1,\cdots,x_{d-1})-xx^\top$. Therefore, we have 
\begin{align*}
    \langle \nabla^2\psi(x)^{-1}\nabla U(x), \nabla U(x) \rangle & = \langle \mathrm{diag}(x)\nabla U(x), \nabla U(x)\rangle - \langle (xx^\top)\nabla U(x), \nabla U(x)\rangle  \\
    &  = \langle \mathrm{diag}(x)\nabla U(x), \nabla U(x)\rangle - \langle x(x^\top\nabla U(x)), \nabla U(x)\rangle  \\
    & = \langle \mathrm{diag}(x)\nabla U(x), \nabla U(x)\rangle - (x^\top \nabla U(x))^2.
\end{align*}
Note that 
\begin{align}
    \langle \mathrm{diag}(x)\nabla U(x), \nabla U(x)\rangle & =  \sum_{i=1}^{d-1} (\mathrm{diag}(x))_{ii} (\partial_i U(x))^2 \nonumber \\
    & = \sum_{i=1}^{d-1} x_i\left( -\frac{n_i+\alpha_i-1}{x_i} + \frac{n_d+\alpha_d-1}{x_d} \right)^2 \nonumber \\
    & = \sum_{i=1}^{d-1} \left[ \frac{(n_i+\alpha_i-1)^2}{x_i} -2 \frac{(n_i+\alpha_i-1)(n_d+\alpha_d-1)}{x_d} + \frac{(n_d+\alpha_d-1)^2}{x_d^2}x_i\right], \label{eq:rel-lip1}
\end{align}
and 
\begin{align}
    (x^\top \nabla U(x))^2 & = \left[ -\sum_{i=1}^{d-1}(n_i+\alpha_i-1) + \frac{n_d+\alpha_d-1}{x_d} \sum_{i=1}^{d-1}x_i \right]^2 \nonumber \\
    & = \left[ -\sum_{i=1}^{d-1}(n_i+\alpha_i-1) + \frac{n_d+\alpha_d-1}{x_d} -(n_d+\alpha_d-1) \right]^2. \label{eq:rel-lip2}
\end{align}
Combining \eqref{eq:rel-lip1} and \eqref{eq:rel-lip2}, we deduce that 
\begin{align*}
    \|\nabla U(x)\|^2_{\nabla^2\psi(x)^{-1}} & = \sum_{i=1}^{d-1} \left[ \frac{(n_i+\alpha_i-1)^2}{x_i} -2 \frac{(n_i+\alpha_i-1)(n_d+\alpha_d-1)}{x_d} + \frac{(n_d+\alpha_d-1)^2}{x_d^2}x_i\right] \\
    & - \left[ -\sum_{i=1}^{d-1}(n_i+\alpha_i-1) + \frac{n_d+\alpha_d-1}{x_d} -(n_d+\alpha_d-1) \right]^2 
\end{align*}
which is unbounded for $x_i \in (0,1), \ \ i=1,2,\cdots,d-1$. \\
\textbf{Computation of $L^S_V$.} We aim to find $L^S_V$ such that $\langle \nabla^2 U(x)v,v \rangle \leq L^S_V \langle \nabla^2 \psi(x)v,v \rangle $ for all $v \in \mathbb{R}^{d-1}$ where, 
\begin{align*}
    & \langle \nabla^2 U(x)v,v \rangle = \sum_{i=1}^{d-1} \frac{v_i^2(n_i+\alpha_i-1)}{x_i^2} + \frac{(n_i+\alpha_i-1)}{x_d^2} \left( \sum_{i=1}^{d-1} v_i \right)^2 \\
    & \langle \nabla^2 \psi(x)v,v \rangle = \sum_{i=1}^{d-1} \frac{v_i^2}{x_i} + \frac{1}{x_d} \left( \sum_{i=1}^{d-1} v_i \right)^2. 
\end{align*} 
If any of the $x_i \to 0$ for all $i=1,2, \dots, d-1$, the quantity 
\begin{equation*}
    \frac{\langle \nabla^2 U(x)v,v \rangle}{\langle \nabla^2 \psi(x)v,v \rangle} = \frac{\sum_{i=1}^{d-1} \frac{v_i^2(n_i+\alpha_i-1)}{x_i^2} + \frac{(n_i+\alpha_i-1)}{x_d^2} \left( \sum_{i=1}^{d-1} v_i \right)^2}{\sum_{i=1}^{d-1} \frac{v_i^2}{x_i} + \frac{1}{x_d} \left( \sum_{i=1}^{d-1} v_i \right)^2}
\end{equation*}
is unbounded and $L^S_V = +\infty.$

\subsection{Calculation of $L_V$ for Section \ref{sec:qgp}.}\label{app:qgp}Because $\mathcal{M} = \prod_{i=1}^n (a_i,b_i)$ may include infinite bounds, we consider the barrier function $\psi$ defined for any $x \in \mathcal{M}$ by $\psi(x) = \sum_{i=1}^n \psi_i(x_i)$ with 
\begin{equation*}
    \psi_i(x_i) = \begin{cases}
        -\log(x_i-a_i) - \log(b_i-x_i) & \text{if} \ a_i, b_i < \infty \\
        \frac{1}{2}x_i^2 -\log(x_i-a_i) & \text{if} \ a_i < \infty, \ b_i=+\infty \\
        \frac{1}{2}x_i^2 -\log(b_i-x_i) & \text{if} \ a_i = -\infty, \ b_i < \infty. 
    \end{cases}
\end{equation*}
Using the derivation in Section \ref{appendix_trunc_gauss}, $\nabla^2 \psi^*$ is the diagonal matrix with 
\begin{equation*}
    \nabla^2\psi^*(\zeta)_{ii} = \begin{cases}
        \frac{(b_i-a_i)^2}{s(\zeta_i)(2+s(\zeta_i))} & \text{if} \ a_i, b_i < \infty \\
        \frac{1}{2}+ \frac{(\zeta_i-a_i)}{2\sqrt{(\zeta_i-a_i)^2+4}} & \text{if} \ a_i < \infty, \ b_i=+\infty \\
        \frac{1}{2} + \frac{(\zeta_i-b_i)}{2\sqrt{(\zeta_i-b_i)^2+4}} & \text{if} \ a_i = -\infty, \ b_i < \infty. 
    \end{cases}
\end{equation*}
We deduce from Sections \ref{appendix:gamma} and \ref{appendix_trunc_gauss} that $V_2(\zeta) = - \log \det \nabla^2 \psi^*(\zeta) $ is $L_{V_2}$-gradient Lipschitz with 
\begin{equation*}
    L_{V_2} = \max \left( \frac{3}{8} \max_{i \in \mathbb{I}} (b_i-a_i)^2, 1 \right)
\end{equation*}
where $\mathbb{I} = \{i \in \{1,\dots,n\}: a_i, b_i < \infty \}.$ For the term $V_1(\zeta) = U \circ \nabla \psi^*(\zeta)$, we proceed as in Section \ref{appendix_trunc_gauss} to write 
\begin{equation*}
    \|\nabla V_1(\zeta_1)-\nabla V_1(\zeta_2)\|_2 \leq AB + CD 
\end{equation*}
with $\displaystyle \|\nabla^2 \psi^*(\zeta_1)\|_2 \leq \max\left( \max_{i \in \mathbb{I}} \frac{3}{8}(b_i-a_i)^2, 1\right)=: c_A.$ The term $B$ is given by 
\begin{equation*}
    B = \| \Sigma^{-1}(\nabla\psi^*(\zeta_1) - \nabla\psi^*(\zeta_2)) \|_2 \leq \|\Sigma^{-1}\|_2 c_A \|\zeta_1-\zeta_2\|_2.
\end{equation*}
The term $C = \|\Sigma^{-1}\nabla\psi^*(\zeta_1)\|_2 \leq \|\Sigma^{-1}\|_2 \|\nabla\psi^*(\zeta_1)\|_2$ is not globally bounded. For the final term $D$, we have 
\begin{align*}
    D & = \|\nabla^2\psi^*(\zeta_1)- \nabla^2\psi^*(\zeta_2)\|_2 \\
    & \leq \max \left( \max_{i \in \mathbb{I}} \frac{(b_i-a_i)^3}{8}, \frac{1}{4} \right) \|\zeta_1-\zeta_2\|_2. 
\end{align*}
Because $C$ is not globally bounded, we proceed to a manual search to ensure that the bound for MZZS is not violated for our experiment.  

\subsection{ Calculating constants for Section \ref{MZZ-LogReg}} \label{app:const_logistic_details} \textbf{Computation of $L_V$.} Following the computations in Section \ref{appendix_trunc_gauss}, we deduce that $V_2(\zeta) = -\log\det \nabla^2 \psi^*(\zeta)$ is $L_{V_2}$-gradient Lipschitz with $L_{V_2}= \frac{12a^2}{8}$. Moreover, for the term $V_1(\zeta) = U \circ \nabla \psi^*(\zeta)$ we proceed again as in Section \ref{appendix_trunc_gauss} to show that 
\begin{equation*}
    \|\nabla V_1(\zeta_1)- \nabla V_1(\zeta_2)\|_2 \leq AB + CD 
\end{equation*}
with $A = \|\nabla^2 \psi^*(\zeta_1)\|_2 \leq a^2/2$.  
\begin{align*}
   B = \|\nabla U(\nabla\psi^*(\zeta_1)) - \nabla U (\nabla\psi^*(\zeta_2)) \|_2 & \leq \frac{1}{4} \sum_{j=1}^n \|x_j\|_2^2 \ \|\nabla \psi^*(\zeta_1) - \nabla\psi^*(\zeta_2) \|_2 \\
    & \leq \frac{1}{4} \sum_{j=1}^n \|x_j\|_2^2 \frac{a^2}{2}\|\zeta_1-\zeta_2\|_2 \\
    & = \frac{a^2}{8} \sum_{j=1}^n \|x_j\|_2^2 \|\zeta_1-\zeta_2\|_2 
\end{align*}
where we have use the fact that $\|\nabla^2\psi^*(\zeta)\|_2 \leq 1/2$. 
For a bound on the term $C$, we have 
\begin{align*}
   C = \| \nabla U(\nabla \psi^*(\zeta)) \|_2 & = \left\| \sum_{j=1}^n x_j \left\{ \frac{\exp(\nabla\psi^*(\zeta)^\top x_j)}{1+ \exp(\nabla\psi^*(\zeta)^\top x_j)}-y_j \right\} \right\|_2 \\
    & \leq \sum_{j=1}^n \|x_j\|_2 \underbrace{\left| \left\{ \frac{\exp(\nabla\psi^*(\zeta)^\top x_j)}{1+ \exp(\nabla\psi^*(\zeta)^\top x_j)}-y_j \right\} \right|}_{\leq 1} \\
   & \leq \sum_{j=1}^n \|x_j\|_2.
\end{align*}
For the final term $D$, we have 
\begin{align*}
    D & = \|\nabla^2\psi^*(\zeta_1)- \nabla^2\psi^*(\zeta_2)\|_2 \\
    & \leq \sup_{1 \leq i \leq d} |f_i(\zeta_i)| \ \|\zeta_1-\zeta_2\|_2 \leq 0.27 a^3 \|\zeta_1-\zeta_2\|_2
\end{align*}
where the function $f_i: \zeta_i \mapsto f_i(\zeta_i)$ is the derivative of the $i$-th diagonal element of $\nabla^2 \psi^*(\zeta)$ with respect to $\zeta_i$. We deduce that $V_1$ is $L_{V_1}$-gradient Lipschitz with 
\begin{equation*}
    L_{V_1} = \frac{a^4}{16} \sum_{j=1}^n \|x_j\|^2_2 + 0.27a^3 \sum_{j=1}^n \|x_j\|_2.
\end{equation*}
Therefore, the potential $V$ is $L_V$-gradient Lipschitz with 
\begin{equation*}
    L_V = \frac{a^4}{16} \sum_{j=1}^n \|x_j\|^2_2 + 0.27a^3 \sum_{j=1}^n \|x_j\|_2 + \frac{12a^2}{8}.
\end{equation*} \\
\textbf{Computation of $M_{\psi}$.} Similarly to Section \ref{appendix_trunc_gauss} we deduce that $M_\psi = 1$. \\
\textbf{Computation of $L^L_V$.} Using the fact that $$\nabla^2 \psi(\theta)^{-1} = \text{diag}\left( \frac{(a^2-\theta^2)^2}{2(a^2+\theta^2)} \right), $$ we deduce that 
\begin{align*}
    \|\nabla U(x)\|^2_{\nabla^2\psi(x)^{-1}} & = \sum_{i=1}^d \frac{(a^2-\theta_i^2)^2}{2(a^2+\theta_i^2)} \left[ \sum_{j=1}^n (x^i_j) \left\{ \frac{\exp(\theta^\top x_j)}{1+ \exp(\theta^\top x_j)}-y_j \right\} \right]^2 \\
    & \leq \frac{a^2}{2} \sum_{i=1}^d \left( \sum_{j=1}^n (x^i_j) \right)^2 =: (L^L_V)^2.
\end{align*} \\
\textbf{Computation of $L^S_V$.} From the expression of $U$, we have for all $v \in \mathbb{R}^d$ 
\begin{align*}
   \langle \nabla^2 U(\theta)v,v \rangle & = \left\langle \sum_{i=1}^d  \left\{ \frac{\exp(\theta^\top x_j)}{1+ \exp(\theta^\top x_j)} \right\}^2 (x_j)(x_j)^\top v, v \right\rangle \\
   & \leq \frac{1}{4} \sum_{j=1}^n \|x_j\|^2 \|v\|^2. 
\end{align*}
Additionally, we have 
\begin{align*}
    \langle \nabla^2 \psi(\theta)v,v \rangle & = \sum_{i=1}^d \frac{2(a^2+\theta_i^2)}{(a^2-\theta_i^2)^2} v_i^2 \\
    & \geq \min_{\theta_i \in [-1,1]} \left[ \frac{2(a^2+\theta_i^2)}{(a^2-\theta_i^2)^2} \right] \|v\|^2 = 2\|v\|^2.
\end{align*}
Hence $$L^S_V = \frac{a^2}{8} \sum_{j=1}^n \|x_j\|^2. $$

\end{appendices}


\bibliography{sn-bibliography}

\end{document}